\documentclass[twocolumn,preprintnumbers,amsmath,amssymb,pra]{revtex4}

\usepackage{graphicx,amsmath}
\usepackage{dcolumn}
\usepackage{bm}
\usepackage{amssymb}
\usepackage{epstopdf}
\usepackage{color}
\graphicspath{{Figures/}}

\begin{document}

\title{Dynamical theory of single-photon transport
through a qubit chain coupled to a one-dimensional
nanophotonic waveguide}

\begin{abstract}
We study the dynamics of a single-photon pulse travelling through
a linear qubit chain coupled to continuum modes in a
one-dimensional (1D) photonic waveguide.  We derive a
time-dependent dynamical theory for qubit amplitudes and for
transmitted and reflected spectra. We show that the requirement
for the photon-qubit coupling to exist only for positive
frequencies can significantly change the dynamics of the system.
First, it leads to an additional photon-mediated dipole-dipole
interaction between qubits which results in the violation of the
phase coherence between them. Second, the spectral lines of
transmitted and reflected spectra crucially depend on the shape of
the incident pulse and the initial distance between the pulse
center and the first qubit in the chain. We apply our theory to
one-qubit and two-qubit systems. For these two cases, we obtain
the explicit expressions for the qubits' amplitudes and for the
photon radiation spectra as time tends to infinity. For the
incident Gaussian wave packet we calculate the line shapes of
transmitted and reflected photons.

\end{abstract}


\date{\today}

\author{Ya. S. Greenberg}\email{yakovgreenberg@yahoo.com}
\affiliation{Novosibirsk State Technical University, Novosibirsk,
Russia}
\author{O. A. Chuikin} \affiliation{Novosibirsk State
Technical University, Novosibirsk, Russia}

\author{A. A. Shtygashev} \affiliation{Novosibirsk State
Technical University, Novosibirsk, Russia}

\author{A. G. Moiseev} \affiliation{Novosibirsk State
Technical University, Novosibirsk, Russia}

\maketitle

\section{Introduction}

Manipulating the propagation of photons in a one-dimensional
waveguide coupled to an array of two-level atoms (qubits) may have
important applications in quantum devices and quantum information
technologies \cite{Rai2001, Roy2017, Gu2017, Sher2023}.

Quantum bits can be implemented with a variety of quantum systems,
such as trapped ions \cite{Leib2003},  photons
\cite{Wang2018,Wang2016,Verma2021}, and quantum dots
\cite{Sam2018,Khor2020}. In particular, superconducting qubits
\cite{Krantz2019,Kjaer2019} have emerged as one of the leading
candidate for scalable quantum processor architecture.

Transmission of a single photon through an array of two-level
atoms embedded in a 1D open waveguide has been extensively studied
both theoretically \cite{Ruos2017,Lal2013,Chang2012} and
experimentally \cite{Mirho2019,Brehm2021,Loo2014}.

Most theoretical calculations of the transmitted and reflected
photon amplitudes  in a 1D open waveguide with atoms placed inside
have been performed within the framework of the stationary theory
in a configuration space
\cite{Shen2009,Cheng2017,Fang2014,Zheng2013} or by alternative
methods such as those based on Lippmann-Schwinger scattering
theory \cite{Roy2011, Huang2013, Diaz2015},  input-output
formalism \cite{Fan2010, Lal2013, Kii2019}, non-Hermitian
Hamiltonian \cite{Green2015}, and  matrix methods
\cite{Green2021,Tsoi2008}.

Even though the stationary theory of the photon transport provides
a useful guide to what one would expect in real experiment, it
does not allow for a description of the dynamics of a qubit
excitation and the evolution of the scattered photon amplitudes.

In practice, qubits are excited by  photon pulses of finite
duration and finite bandwidth. Therefore, to study the real-time
evolution of  photon transport and atomic excitation the
time-dependent dynamical theory has been developed \cite{Chen2011,
Liao2015, Liao2016a, Liao2016b, Zhou2022}. This theory relies on
two assumptions. The first is the Wigner-Weisskopf approximation
in which the rate of spontaneous emission to the guided mode is
much less than the qubit frequency, $\Gamma(\omega)\ll\Omega$.
Therefore, the decay rate $\Gamma$ is assumed
frequency-independent and is considered at the resonant frequency
$\Omega$, $\Gamma(\omega)=\Gamma(\Omega)\equiv\Gamma$. The second
assumption is more subtle. It is assumed that the photon-mediated
coupling $g(\omega)$ may be extended to negative frequencies which
allows to move the lower bound of some frequency integrals to
minus infinity. In this case, the transmitted and reflected photon
amplitudes become proportional to the spectral density of incident
pulse, $\gamma_0(\omega)$ \cite{Liao2015}. It is believed that the
continuation of $g(\omega)$ to negative frequencies is justified
beyond rotating wave approximation (RWA) by accounting for counter
rotating terms (see Supplement in \cite{Gonz2013}). The extension
to negative frequencies  are generally relies on the assumption
that non-RWA contribution is negligible which is not always the
case. Moreover, it is not justified from physical arguments:
because the continuum starts at $\omega=0 $, the density of states
is zero for $\omega<0$. Therefore, $g(\omega)= 0$ for $\omega<0$,
that is the coupling does not exist at negative frequencies
\cite{Cohen2004}. Thus, its continuation to negative frequencies
makes no sense. From the other hand, we may consider this
continuation as a pure mathematical trick which can be justified
if the negative frequency integral provides a small correction.
This is indeed the case if the distance between qubits $d$ exceeds
the qubit wavelength $\lambda=2\pi v_g/\Omega$. However, if
$d<\lambda$ the discrepancy can be significant(see Appendix B in
\cite{Green2021a}).

Another consequence of "no continuation" of $g(\omega)$ to
negative frequencies is more complicated dependence of photon
radiation on the shape of incident pulse: the transmitted and
reflected photon amplitudes are no longer proportional to
$\gamma_0(\omega)$.

From the point of view of device applications, a control pulse
generator should be placed as close as possible to the measured
qubit circuitry, for example, in the same chip with
superconducting qubit at millikelvin temperatures \cite{Lec2021}.
We show that such an arrangement leads to significant
modifications of the photon radiation spectra if the distance
between the initial position of the peak of an incident pulse and
the first qubit in a chain is comparable to or less than the pulse
width in configuration space.

The paper is organized as follows. In Section II we briefly
describe our system consisting of N identical qubits in a one-
dimensional infinite waveguide. The system is described by
Jaynes-Cummings Hamiltonian in rotating wave approximation (RWA).
The Hilbert space is restricted to the single-excitation subspace.

In Section III the general time-dependent equations for qubits'
amplitudes, $\beta_n(t)$ and photon forward, $\gamma(\omega,t)$
and backward $\delta(\omega,t)$ radiation spectra for a
single-photon pulse scattering from an array of $N$ identical
qubits are derived. These equations are then solved by application
of Heitler's method, which allows us to find the desired solutions
for time-dependent amplitudes. The main result of this section is
the modification of the equations for qubits amplitudes which
appears due to non continuation of the interqubit interaction to
negative frequencies (see Eq.\ref{23}). First, it leads to the
additional photon-mediated dipole-dipole interaction between
qubits which results in the violation of the phase coherence
between them. Second, the spectral lines of transmitted and
reflected spectra crucially depend on the shape of incident pulse
and on the initial distance between the pulse center and first
qubit in the chain.

The theoretical results obtained in Section III are applied for
the calculations of photon forward and backward radiation  spectra
for scattering  of a moving Gaussian pulse from a single qubit
(Section IV) and from two-qubit system (Section V).  We show that
the form of radiation spectra crucially depends on the distance
between the initial position of Gaussian peak and the first qubit
in a chain. The radiation spectra are substantially modified if
this distance is comparable to or less than the pulse width in a
configuration space.

Summary of our work is presented in the conclusion (Section VI).
Some additional technical calculations are given in Appendixes.

\section{The model}
We consider a system consisting of N identical qubits in a one-
dimensional infinite waveguide. In the continuum mode
representation this system can be described by a Jaynes-Cummings
Hamiltonian which accounts for the interaction between qubits and
electromagnetic field \cite{Blow1990,Dom2002} (from now on we use
the units $\hbar = 1$ throughout the paper, therefore, all
energies are expressed in frequency units):
  \begin{equation}\label{1}
\begin{gathered}
  H = \frac{1}
{2} \sum\limits_{n = 1}^N {} \left( {1 + \sigma _Z^{(n} } \right)\Omega  \hfill \\
   + \int\limits_0^\infty  {} \omega a^\dag (\omega) a(\omega ) d\omega  + \int\limits_0^\infty  {} \omega b^\dag  (\omega) b(\omega) d\omega  \hfill \\
   + \int\limits_0^\infty  {} a^\dag (\omega) S_-  (\omega) d\omega  + \int\limits_0^\infty  {} a(\omega) S_+  (\omega) d\omega  \hfill \\
   + \int\limits_0^\infty  {} b^\dag  (\omega )S_- ^* (\omega )d\omega  + \int\limits_0^\infty  {} b(\omega )S_ + ^* (\omega )d\omega,  \hfill \\
\end{gathered}
  \end{equation}
where $\sigma^{(n)}_z$ is a Pauli spin operator, $\Omega$ is a
resonant frequency of $n$th qubit, $\omega$ is a photon frequency.

The photon creation and annihilation operators $a^\dag(\omega)$,
$a(\omega)$, and $b^\dag(\omega)$, $b(\omega)$ describe forward and
backward scattering waves, respectively. They are independent of
each other and satisfy the usual continuous-mode commutation
relations \cite{Blow1990}:
  \begin{equation}\label{comut}
\left[ {a(\omega ),a^\dag  (\omega ')} \right] = \left[ {b(\omega ),b^\dag (\omega ')} \right] = \delta(\omega - \omega ').
  \end{equation}

In (\ref{1}) we introduced collective atomic spin operators:
  \begin{equation}\label{2}
\begin{gathered}
  S_ -  (\omega ) = \sum\limits_{n = 1}^N {} g(\omega )e^{ - ikx_n } \sigma _ - ^{(n)} , \hfill \\
  S_ +  (\omega ) = \sum\limits_{n = 1}^N {} g(\omega )e^{ikx_n } \sigma _ + ^{(n)}.  \hfill \\
\end{gathered}
  \end{equation}
where $k=\omega/v_g$, $v_g$ is the group velocity of
electromagnetic waves, $x_n$ is a spatial coordinate of the $n$th
qubit. $\sigma_-^{(n)} = {\left|g\right\rangle_{nn}} \left\langle
e \right|$ and $\sigma _ + ^{(n)} = {\left| e \right\rangle _{nn}}
\left\langle g \right|$ are the lowering and raising atomic
operators which lower or raise a state of the $n$th qubit. A spin
operator $\sigma_z^{(n)}=|e\rangle_{nn}\langle
e|-|g\rangle_{nn}\langle g|$. The quantity $g(\omega)$ in
(\ref{1}) is the coupling between qubit and the photon field in a
waveguide \cite{Dom2002}:
  \begin{equation}\label{g}
g(\omega ) = \sqrt {\frac{{\omega d^2 }} {{4\pi \varepsilon _0
\hbar v_g S}}},
  \end{equation}
where $d$ is the off diagonal matrix element of a dipole operator,
$S$ is the effective transverse cross section of the modes in
one-dimensional waveguide, $v_g$ is the group velocity of
electromagnetic waves. We assume that the coupling is the same for
forward and backward waves.

Note that the dimension of the coupling constant $g(\omega)$ is
not a frequency, $\omega$ but a square of frequency,
$\sqrt{\omega}$, and, as it follows from (\ref{comut}), the
dimension of creation and annihilation operators is
$1/\sqrt{\omega}$.

Below we consider a single-excitation subspace with either a
single photon is in a waveguide and all qubits are in the ground
state, or there are no photons in a waveguide with the only $n$th
qubit in the chain being excited. Therefore, we truncate Hilbert
space to the following states:
  \begin{equation}\label{H-space}
\begin{array}{l}
 \left| {G,1_k } \right\rangle  = \left| {g_1 ,g_2,....g_N } \right\rangle  \otimes \left| {1_k } \right\rangle ;
 \\\\
 \left| {n,0_k } \right\rangle  = \left| {g_1, g_2....g_{n-1},e_n, g_{n+1}..... g_N } \right\rangle  \otimes \left| {0_k } \right\rangle ;
\end{array}
  \end{equation}

The Hamiltonian (\ref{1}) preserves the number of excitations
(number of excited qubits + number of photons).
In our case the number of excitations is equal to one.
Therefore, at any instant of time the system will remain
within a single-excitation subspace.

The trial wave function of an arbitrary single-excitation state
can then be written in the form:
  \begin{equation}\label{3}
\begin{gathered}
  \Psi (t) = \sum\limits_{n = 1}^N {} \beta_n (t)e^{-i\Omega t} \left| {n,0} \right\rangle  \hfill \\
   + \int\limits_0^\infty  {d\omega \gamma(\omega ,t) e^{-i\omega t} a^\dag (\omega )\left| {G,0} \right\rangle }  \hfill \\
   + \int\limits_0^\infty  {d\omega \delta (\omega ,t) e^{-i\omega t} b^\dag (\omega )\left| {G,0} \right\rangle },  \hfill \\
\end{gathered}
  \end{equation}
where $\beta_n(t)$  is the amplitude of $n$th qubit,
$\gamma(\omega,t)$ and $\delta(\omega,t)$ are single-photon
amplitudes which are related to a spectral density of forward and
backward radiation, respectively.

\section{Equations for qubits' and photon amplitudes}
The Schrodinger equation $ i\frac{{d\Psi (t)}} {{dt}} = H\Psi (t)$
yields the equations for the amplitudes:
  \begin{equation}\label{5}
\begin{gathered}
  \frac{{d\beta _n }}
{{dt}} =  - i\int\limits_0^\infty  {} d\omega g(\omega )\gamma (\omega ,t)e^{ikx_n } e^{ - i(\omega  - \Omega )t}  \hfill \\
   - i\int\limits_0^\infty  {} d\omega g(\omega )\delta (\omega ,t)e^{ - ikx_n } e^{ - i(\omega  - \Omega )t},  \hfill \\
\end{gathered}
  \end{equation}
  \begin{equation}\label{6}
\frac{{d\gamma (\omega ,t)}} {{dt}} =  - ig(\omega )e^{i(\omega  -
\Omega )t} \sum\limits_{n = 1}^N {} \beta _n (t)e^{ - ikx_n },
  \end{equation}
  \begin{equation}\label{7}
\frac{{d\delta (\omega ,t)}} {{dt}} =  - ig(\omega )e^{i(\omega  -
\Omega )t} \sum\limits_{n = 1}^N {} \beta _n (t)e^{ikx_n }.
  \end{equation}

The system equations (\ref{5}), (\ref{6}), and (\ref{7}) have a
physical sense only for $t>0$. Therefore they should be
supplemented by the initial conditions:

\begin{equation}\label{cond1}
    \beta_n(0)=0;\,\delta(\omega,0)=0;\,\gamma(\omega,+0)=\gamma_0(\omega)
\end{equation}
where $\gamma_0(\omega)$ is the initial incident pulse which for a
single scattering photon is assumed to be normalized to unity,
$\int\limits_0^\infty  {d\omega } \left| {\gamma _0 (\omega )}
\right|^2=1$.

For the analytical reasons it is convenient to extend, by the
Heitler method \cite{Heit1954}, the solution of system (\ref{5}),
(\ref{6}), and (\ref{7}) to the negative time semi axis, assuming
that

\begin{equation}\label{cond2}
    \beta_n(t)=0;\,\delta(\omega,t)=0;\,\gamma(\omega,t)=0;\,\,t<0
\end{equation}
Therefore, the amplitudes $\beta_n(t)$ and $\delta(\omega,t)$ are
continuous over the entire time axis, while the function
$\gamma(\omega,t)$ has a jump at $t = 0$. We will take this into
account by adding the inhomogeneous term into the equation
(\ref{6}):

\begin{equation}\label{6a}
\frac{{d\gamma (\omega ,t)}} {{dt}} =  - ig(\omega )e^{i(\omega  -
\Omega )t} \sum\limits_{n = 1}^N {} \beta _n (t)e^{ - ikx_n
}+\gamma_0(\omega)\delta(t),
  \end{equation}
The inhomogeneous term in (\ref{6a}) provides the jump of the norm
of the wavefunction (\ref{3})
 \begin{equation}\label{norm}
\sum\limits_{n = 1}^N {\left| {\beta _n (t)} \right|^2 }  +
  \int\limits_0^\infty  {d\omega } \left| {\gamma (\omega ,t)}
  \right|^2  + \int\limits_0^\infty  {d\omega } \left| {\delta (\omega ,t)}
  \right|^2 ,
  \end{equation}
from zero at $t<0$ to unity at $t>0$.

We note that as $t\rightarrow\infty$ the qubit amplitude
$\beta_n(t)\rightarrow 0$, therefore the norm (\ref{norm}) reduces
to:
  \begin{equation}\label{4a}
\int\limits_0^\infty  {d\omega } \left| {\gamma (\omega
,t\rightarrow\infty)}
  \right|^2  + \int\limits_0^\infty  {d\omega } \left| {\delta (\omega ,t\rightarrow\infty)}
  \right|^2=1.
  \end{equation}

Therefore, the system equations (\ref{5}), (\ref{6a}), and
(\ref{7}) are now defined for all values of $t$, negative and
positive.

In order to solve the equations (\ref{5}), (\ref{6a}), and
(\ref{7}) we apply the Fourier transform to the amplitudes
$\beta_n(t), \gamma(\omega,t), \delta(\omega,t)$:
\begin{equation}\label{14}
\beta _n (t) = \int\limits_{ - \infty }^\infty  {\frac{{d\nu }}
{{2\pi }}} \beta _n (\nu )e^{ - i(\nu  - \Omega )t},
  \end{equation}
  \begin{equation}\label{15}
\gamma (\omega ,t) = \int\limits_{ - \infty }^\infty  {\frac{{d\nu
}} {{2\pi }}} \gamma (\omega ,\nu )e^{ - i(\nu  - \omega )t},
  \end{equation}
  \begin{equation}\label{16}
\delta (\omega ,t) = \int\limits_{ - \infty }^\infty  {\frac{{d\nu
}} {{2\pi }}} \delta (\omega ,\nu )e^{ - i(\nu  - \omega )t}.
  \end{equation}

From (\ref{5}), (\ref{6a}) and (\ref{7}) we obtain:
\begin{equation}\label{10}
\begin{gathered}
    - i(\nu  - \Omega )\beta _n (\nu ) =  - i\int\limits_0^\infty  {} d\omega g(\omega )\gamma (\omega ,\nu )e^{ikx_n }  \hfill \\
   - i\int\limits_0^\infty  {} d\omega g(\omega )\delta (\omega ,\nu )e^{ - ikx_n },  \hfill \\
\end{gathered}
  \end{equation}

  \begin{equation}\label{17}
 - i(\nu  - \omega )\gamma (\omega ,\nu ) = -
ig(\omega )\sum\limits_{n = 1}^N {} \beta _n (\nu )e^{ - ikx_n }+
\gamma _0 (\omega ),
  \end{equation}
  \begin{equation}\label{18}
- i(\nu  - \omega )\delta (\omega ,\nu ) =  - ig(\omega
)\sum\limits_{n = 1}^N {} \beta _n (\nu )e^{ikx_n }.
  \end{equation}

Because both frequency variables $\omega$ and $\nu$ belong to the
continuum we will solve the equations (\ref{17}) and (\ref{18})
using  the general approach first developed in Heitler's papers
and considered in detail in  \cite{Heit1954}, and subsequently
used  for the study of light scattering by a dense ensemble of
atoms \cite{Sok2011,Kur2015, Kur2016}:
  \begin{equation}\label{19}
\gamma (\omega ,\nu ) = g(\omega )\sum\limits_{n = 1}^N {} \beta
_n (\nu )e^{ - ikx_n } \zeta (\nu  - \omega ) + i\gamma _0 (\omega
)\zeta (\nu  - \omega ),
  \end{equation}
  \begin{equation}\label{20}
\delta (\omega ,\nu ) = g(\omega )\sum\limits_{n = 1}^N {} \beta
_n (\nu )e^{ikx_n } \zeta (\nu  - \omega ),
  \end{equation}
where we introduce a singular zeta function defined as:
  \begin{equation}\label{21}
\zeta (\nu  - \omega ) =  - i\pi \delta (\nu  - \omega ) +
P\frac{1} {{\nu  - \omega }}.
  \end{equation}
In (\ref{21}), $\delta(\nu-\omega)$ is a Dirac delta-function, and $P$ is a Cauchy principal value.

Next we substitute $\gamma(\omega,\nu)$, $\delta(\omega,\nu)$ in
(\ref{10}) with their expressions (\ref{19}) and (\ref{20}). Thus,
we obtain a set of linear equations which define the Fourier
components of the qubits' amplitudes $\beta_n(\nu)$:
  \begin{equation}\label{22}
\begin{gathered}
  \left( {\nu  - \Omega  - 2\int\limits_0^\infty  {} d\omega g^2 (\omega )\zeta (\nu  - \omega )} \right)\beta _n (\nu ) \hfill \\
   - 2\sum\limits_{n' \ne n}^N {} \beta _{n'} (\nu )\int\limits_0^\infty  {} d\omega g^2 (\omega )\cos \left( {k(x_n  - x_{n'} )} \right)\zeta (\nu  - \omega ) \hfill \\
   = i\int\limits_0^\infty  {} d\omega g(\omega )\gamma _0 (\omega )e^{ikx_n } \zeta (\nu  - \omega ).  \hfill \\
\end{gathered}
  \end{equation}

Using (\ref{21}) in (\ref{22}) we obtain:
  \begin{equation}\label{23}
\begin{gathered}
  \left( {\nu  - \Omega  - F(\nu ) + i\frac{{\Gamma (\nu )}}
{2}} \right)\beta _n (\nu ) \hfill \\
   + i\frac{{\Gamma (\nu )}}
{2}\sum\limits_{n' \ne n}^N {} \beta _{n'} (\nu )\left( {e^{ik_\nu  \left| {x_n  - x_{n'} } \right|}  + iG(k_\nu  \left( {x_n  - x_{n'} } \right))} \right) \hfill \\
   = i\int\limits_0^\infty  {} d\omega g(\omega )\gamma _0 (\omega )e^{ikx_n } \zeta (\nu  - \omega ),  \hfill \\
\end{gathered}
  \end{equation}
where (the derivation is given in Appendix A):
  \begin{equation}\label{G}
\begin{gathered}
  G(k_\nu  d_{nn'} ) =  - \frac{4}
{{\Gamma (\nu )}}\int\limits_0^\infty  {d\omega \frac{{g^2 (\omega
)\cos k_\nu  d_{nn'} }}
{{\omega  + \nu }}}  \hfill \\
   = \frac{1}
{\pi }\cos k_\nu  d_{nn'} \text{Ci}(\left| {k_\nu  d_{nn'} }
\right|) \\+ \frac{1} {\pi }\sin k_\nu  d_{nn'} \left( { -
\frac{\pi }
{2}\operatorname{sgn} (d_{nn'} ) + \text{Si}(k_\nu  d_{nn'} )} \right), \hfill \\
\end{gathered}
  \end{equation}
$d_{nn'}=x_n-x_{n'}$, $\Gamma(\nu)=4\pi g^2(\nu)$, $F(\nu ) =
2P\int\limits_0^\infty  {\frac{{g^2 (\omega )}} {{\nu  - \omega
}}} d\omega$, $k_{\nu}=\nu/v_g$, $k=\omega/v_g$; $\text{Ci}(x)$
and $\text{Si}(x)$ are the cosine and sine integral functions:
  \begin{equation}\label{CS}
{\text{Ci}}(x) =  - \int\limits_x^\infty  {dt\frac{{\cos t}} {t}}
,\;{\text{    Si}}(x) = \int\limits_0^x {dt\frac{{\sin t}} {t}}.
  \end{equation}
Because $\text{Ci}(x)$ is defined for $x>0$ and
$\text{Si}(-x)=-\text{Si}(x)$, the quantity G(x) is the even
function, $G(-x)=G(x)$.

It is worthy noting that the equations (\ref{23}) differ from
conventional ones in that they contain the quantity $G(kd_{nn'})$
and Cauchy principal part integral in the right hand side in
(\ref{23}). This modification of the equations for qubits
amplitudes appears due to non continuation of the interqubit
interaction to negative frequencies.

As the quantity $G(kd_{nn'})$ is a real function it modifies the
interaction between qubits giving rise to the shift of their
frequencies. In addition, it violates the phase coherence because
it is not possible to switch off the interaction between identical
qubits simply by taking $kd=n\pi$, where $n$ is integer.
 We consider the properties of $G(kd)$ in more details in Section V for two-qubit system.

From a set of linear equations (\ref{23}) we can find $N$ Fourier
amplitudes $\beta_n(\nu)$. The next step is to find from
(\ref{19}) and (\ref{20}) photon Fourier amplitudes
$\gamma(\omega,\nu)$ and $\delta(\omega,\nu)$. Finally, from
Fourier transform (\ref{14}), (\ref{15}) and (\ref{16}) we find
the qubits' and photon amplitudes in a time domain.

However, if we are interested in the scattering amplitudes for
$t\rightarrow\infty$ a simpler way is to start from a formal
solution of equations (\ref{6a}) and (\ref{7}):

 \begin{align}\label{8}
\gamma (\omega ,t) = \gamma _0 (\omega ) - i\sum\limits_{n = 1}^N
{} \int\limits_{-\infty}^t {d\tau } \beta _n (\tau )g(\omega
)e^{i(\omega - \Omega )\tau } e^{ - ikx_n },
 \end{align}
 \begin{align}\label{9}
\delta (\omega ,t) =  - i\sum\limits_{n = 1}^N {}
\int\limits_{-\infty}^t {d\tau } \beta _n (\tau )g(\omega
)e^{i(\omega - \Omega )\tau } e^{ikx_n }.
 \end{align}

If we set the upper limit of the integrals (\ref{8}), (\ref{9}) to
$+\infty$ we obtain:
  \begin{equation}\label{24}
\gamma (\omega ,t \to \infty ) = \gamma _0 (\omega ) - ig(\omega
)\sum\limits_{n = 1}^N {} \beta _n (\omega )e^{ - ikx_n },
  \end{equation}
  \begin{equation}\label{25}
\delta (\omega ,t\to \infty) =  - ig(\omega )\sum\limits_{n = 1}^N
{} \beta _n (\omega )e^{ikx_n }.
  \end{equation}
where $\beta_n(\omega)$ is a solution of equation (\ref{23}) and
is formally given by the inverse Fourier transform of equation
(\ref{14}):
\begin{equation}\label{11}
\beta _n (\omega ) = \int\limits_{-\infty}^\infty  {dt} \beta _n
(t)e^{i(\omega - \Omega )t},
  \end{equation}

 Below we will show the application of this
method to the cases of one and two qubits in a waveguide for which
the analytical solutions can be obtained in a closed form.

\section{Single qubit}
For a single two-level atom located at $x=0$ the solution of
equation (\ref{23}) is given by:

\begin{equation}\label{26}
\beta (\omega ) = \frac{{i\int\limits_0^\infty  {} d\omega
'g(\omega ')\gamma _0 (\omega ')\zeta (\omega  - \omega ') }}
{{\left( {\omega  - \Omega  - F(\omega ) + i\frac{{\Gamma (\omega
)}} {2}} \right)}}.
\end{equation}

With the aid of (\ref{24}) and (\ref{25}) we calculate the
transmitted and reflected photon amplitudes
$\gamma(\omega,t\rightarrow\infty)$ and
$\delta(\omega,t\rightarrow\infty)$:

\begin{equation}\label{27}\begin{gathered}
  \gamma (\omega ,t \to \infty ) = \gamma _0 (\omega )\frac{{\omega  - \Omega  - F(\omega ) + i\frac{{\Gamma (\omega )}}
{4}}} {{\omega  - \Omega  - F(\omega ) + i\frac{{\Gamma (\omega
)}}
{2}}} \hfill \\
   + \frac{{g(\omega )P\int\limits_0^\infty  {\frac{{d\omega 'g(\omega ')\gamma _0 (\omega ')}}
{{\omega  - \omega '}}} }} {{\omega  - \Omega - F(\omega ) +
i\frac{{\Gamma (\omega )}}
{2}}}, \hfill \\
\end{gathered}
\end{equation}

\begin{equation}\label{28}
\begin{gathered}
  \delta (\omega ,t \to \infty ) = \gamma _0 (\omega )\frac{{ - i\frac{{\Gamma (\omega )}}
{4}}} {{\omega  - \Omega  - F(\omega ) + i\frac{{\Gamma (\omega
)}}
{2}}} \hfill \\
   + \frac{{g(\omega )P\int\limits_0^\infty  {\frac{{d\omega 'g(\omega ')\gamma _0 (\omega ')}}
{{\omega  - \omega '}}}  }} {{\omega  - \Omega - F(\omega ) +
i\frac{{\Gamma (\omega )}}
{2}}}. \hfill \\
\end{gathered}
\end{equation}
As it is seen from  $\gamma (\omega ,t \to \infty ) - \delta
(\omega ,t \to \infty ) = \gamma _0 (\omega )$, the condition
which is only valid for the scattering from a single qubit. The
flux conservation is fulfilled not at a single frequency, but as
integral quantity (see normalizing condition (\ref{4a})).

It is worth noting that the second terms in (\ref{27}) and
(\ref{28}) which contain Cauchy principal value integrals are
responsible for the dependence of the scattering spectra on the
initial distance between the center of incident Gaussian pulse
and the first qubit in the chain.

We note that the derivation of the transmitted and reflected
amplitudes (\ref{27}) and (\ref{28}) is not restricted to
Wigner-Weisskopf approximation $\Gamma(\Omega)\ll\Omega$. It is
also valid  for strong coupling where $\Gamma(\Omega)\leq\Omega$.

The evolution of qubit's amplitude $\beta(t)$ is obtained from
(\ref{14}) with $\beta(\omega)$ from (\ref{26}):

\begin{equation}\label{bt}
\beta (t) = \int\limits_{ - \infty }^\infty  {\frac{{d\omega }}
{{2\pi }}} \frac{{i\int\limits_0^\infty  {d\omega 'g(\omega
')\gamma _0 (\omega ')\zeta (\omega  - \omega ')} }} {{\omega  -
\Omega  - F(\omega ) + i\frac{{\Gamma (\omega )}} {2}}}e^{ -
i(\omega  - \Omega )t}.
\end{equation}

In Wigner-Weisskopf approximation we can obtain a simple
analytical expression for the evolution of qubit's amplitude
(\ref{bt}) (see Appendix B):

\begin{equation}\label{bt0}
\beta (t) = \int\limits_0^\infty  {d\omega g(\omega )\gamma _0
(\omega )} \frac{{e^{ - \frac{\Gamma } {2}t}  - e^{ - i(\omega  -
\Omega )t} }} {{\left( {\omega  - \Omega  + i\frac{\Gamma } {2}}
\right)}}.
\end{equation}

The expressions (\ref{27}), (\ref{28}) look rather different from
the transmitted and reflected amplitudes which were found in a
framework of stationary scattering approach for a monochromatic
signal scattered by a two-level atom in an 1D open waveguide
\cite{Shen2005a,Shen2005b}:

\begin{equation}\label{I1}
\gamma(\omega) = \frac{{\omega   - \Omega }} {{\omega   - \Omega +
i\frac{\Gamma}{2} }},
\end{equation}
\begin{equation}\label{I2}
\delta(\omega) = \frac{{ - i\frac{\Gamma}{2} }} {{\left( {\omega
- \Omega + i\frac{\Gamma}{2} } \right)}}.
\end{equation}

Here and below $\Gamma$ is the full width of the spectral lines.

Below we show that in some cases the expressions (\ref{27}) and
(\ref{28}) provide the stationary scattering results.

We assume that the incident wave packet is a narrow pulse which
can be approximated by a delta function:
\begin{equation}\label{29}
\gamma _0 (\omega ) = A\delta (\omega  - \omega _S ) = \frac{A}
{{2\pi }}\int\limits_{ - \infty }^{ + \infty } {d\lambda
e^{i\lambda (\omega  - \omega _S )} }.
\end{equation}

Plugging (\ref{29})  in (\ref{27}) and (\ref{28})  we obtain for
principal value integral:

\begin{equation}\label{30}
\begin{gathered}
  P\int\limits_0^\infty  {} d\omega '\frac{{g(\omega ')\gamma _0 (\omega ')}}
{{\omega  - \omega '}} \approx g(\omega )\frac{A} {{2\pi
}}\int\limits_{ - \infty }^\infty  {d\lambda e^{ - i\lambda \omega
_S } } P\int\limits_{ - \infty }^\infty  {} d\omega
'\frac{{e^{i\lambda \omega '} }}
{{\omega  - \omega '}} \hfill \\
   =  - i\pi g(\omega )\frac{A}
{{2\pi }}\int\limits_{ - \infty }^\infty  {d\lambda e^{ - i\lambda \omega _S } } e^{i\lambda \omega }  =  - i\pi g(\omega )A\delta (\omega  - \omega _S ) \hfill \\
   \equiv  - i\pi g(\omega )\gamma _0 (\omega ). \hfill \\
\end{gathered}
\end{equation}
When deriving (\ref{30}) we first assume that the coupling
$g(\omega)$ is a slow function of $\omega$, so that we take it out
of the integral. Second, we put the lower bound in principal value
integral to minus infinity which allows the application of
Kramers-Kronig relation $P\int\limits_{ - \infty }^{ + \infty }
{\frac{{e^{i\lambda \omega '} }} {{\omega  - \omega '}}d\omega '}
=  - i\pi e^{i\lambda \omega }$.

If we use the result (\ref{30}) in (\ref{27}) and (\ref{28}) and
assume that the atom is initially not excited, we obtain for
transmitted and reflection amplitudes the expressions which are
known from the stationary theories:

\begin{equation}\label{31}
\gamma (\omega ,t \to \infty ) = \gamma _0 (\omega )\frac{{\omega
- \Omega  - F(\omega )}} {{\omega  - \Omega  - F(\omega ) +
i\frac{{\Gamma (\omega )}} {2}}},
\end{equation}

\begin{equation}\label{32}
\delta (\omega ,t \to \infty ) = \gamma _0 (\omega )\frac{{ -
i\frac{{\Gamma (\omega )}} {2}}} {{\omega  - \Omega  - F(\omega )
+ i\frac{{\Gamma (\omega )}} {2}}}.
\end{equation}

These expressions are similar to those obtained in the framework
of time-dependent approach in \cite{Liao2015}, where
$\gamma_0(\omega)$ was arbitrary pulse shape. However, here the
expressions (\ref{31}) and (\ref{32}) are valid if the incident
photon is a delta pulse (\ref{29}). For this case, these
expressions are exact asymptotic solutions for time dependent
scattering of a single-photon pulse from a two-level atom
\cite{Green2023}.

For arbitrary shape of $\gamma_0(\omega)$ the exact expressions
(\ref{27}) and (\ref{28}) must be used.

Below we present several plots for qubit's amplitude, forward and
backward photon spectra calculated from expressions (\ref{bt0}),
(\ref{27}), and (\ref{28}) for incident travelling Gaussian pulse:

\begin{equation}\label{Gauss}
\gamma _0 (\omega ) = \left( {\frac{2} {{\pi \Delta ^2 }}}
\right)^{1/4} \exp \left( {i(\omega  - \omega _s )t_0  -
\frac{{(\omega  - \omega _s )^2 }} {{\Delta ^2 }}} \right),
\end{equation}
where $\Delta$ is the width of Gaussian pulse in the frequency
domain, $\Delta x=v_g/\Delta$ is the width of Gaussian pulse in
space, $t_0=x_0/v_g$ is the time that it takes for the center of a
Gaussian packet to travel from the point $-x_0$ to the point $x=0$
where the qubit is located. For our parameter values $v_g=3\times
10^8$m/s, $\Omega/2\pi=5$ GHz, $\Delta/\Omega=0.1$, the width of
the packet in space $\Delta x=v_g/\Delta\approx 10$ cm.

\begin{figure}
  \includegraphics[width=8cm]{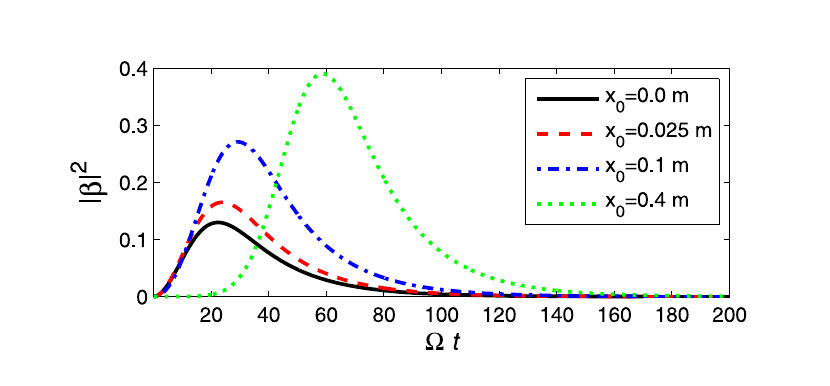}\\
  \caption{Probability of the qubit excitation, calculated from (\ref{bt0}) for different
  distances, $x_0$ of a peak of the incident Gaussian pulse from the qubit.
  $x_0=0$, solid black line; $x_0=0.025$ m, dashed red line; $x_0=0.1$
  m, dashed-dotted blue line; $x_0=0.4$, dotted green line.
$\Delta/\Omega=0.1$, $\gamma=\Gamma/\Omega=0.05$.}\label{Fig5}
\end{figure}

We assume that at the initial time $t=0$ qubit is in its ground
state and the maximum of the envelope of a Gaussian pulse is
located at the distance $x_0$ from the qubit. The plots of qubit's
excitation probability $|\beta|^2$ for four values of initial
distance $x_0$ between Gaussian peak and qubit are shown in Fig.
\ref{Fig5} for $\Delta/\Omega=0.1$ and
$\gamma=\Gamma/\Omega=0.05$. Our calculations show that the
maximum excitation, $|\beta_{max}|^2\approx0.38$, is obtained if
$x_0$ is large compared with the pulse width, $v_g/\Delta$.
Similar result ($|\beta_{max}|^2=0.4$) was obtained in
\cite{Chen2011} where the initial position of the peak of the
pulse was at the distance $10v_g/\Delta$ from the qubit.

As is known, the maximum excitation $|\beta_{max}|^2=0.5$ is
achieved when qubit is illuminated from one side by the stationary
plane wave \cite{Chen2011}. The reason for this is that the
qubit-field interaction Hamiltonian allows the transformation of
the forward and backward propagating continua  to a bright
$B^\dag$ and a dark $D^\dag$ continuum, given by $B^\dag = (a^\dag
+ b^\dag)/\sqrt{2}$, $D^\dag = (a^\dag - b^\dag)/\sqrt{2}$
\cite{Shen2007}. The dark continuum is decoupled from the qubits,
allowing $50\%$ of the incident wave $a^\dag = (B^\dag +
D^\dag)/\sqrt{2})$ transmit through the waveguide unchanged. The
bright component of the incident wave is scattered by the qubits
providing $50\%$ of its maximum excitation. The full inversion
$|\beta|^2=1$ can be obtained with the incident pulse of special
shape \cite{Stob2009, Reph2010}.

We rewrite (\ref{27}) and (\ref{28}) in Wigner-Weisskopf
approximation, $\Gamma(\omega) = \Gamma(\Omega) \equiv\Gamma$,
$g(\omega)=g(\Omega)=(\Gamma/4\pi)^{1/2}$. The frequency shift
$F(\omega)=F(\Omega)$ is incorporated implicitly in $\Omega$:

\begin{equation}\label{27a}
\gamma_{WW} (\omega ) = \gamma _0 (\omega )\frac{{\omega - \Omega
+ i\frac{\Gamma } {4}}} {{\omega  - \Omega  + i\frac{\Gamma }
{2}}} + \frac{{\frac{\Gamma } {{4\pi }}P\int\limits_0^\infty
{\frac{{d\omega '\gamma _0 (\omega ')}} {{\omega  - \omega '}}} }}
{{\omega  - \Omega  + i\frac{\Gamma } {2}}},
\end{equation}

\begin{equation}\label{28a}
\delta _{WW} (\omega ) = \gamma _0 (\omega )\frac{{ -
i\frac{\Gamma } {4}}} {{\omega  - \Omega  + i\frac{\Gamma } {2}}}
+ \frac{{\frac{\Gamma } {{4\pi }}P\int\limits_0^\infty
{\frac{{d\omega '\gamma _0 (\omega ')}} {{\omega  - \omega '}}} }}
{{\omega  - \Omega  + i\frac{\Gamma } {2}}}.
\end{equation}

We compare these expressions with those obtained in
\cite{Liao2015} for arbitrary pulse shape and with the extension
of the coupling to negative frequencies:

\begin{equation}\label{31a}
\gamma (\omega  ) = \gamma _0 (\omega )\frac{{\omega - \Omega  }}
{{\omega  - \Omega   + i\frac{{\Gamma }} {2}}},
\end{equation}

\begin{equation}\label{32a}
\delta (\omega ) = \gamma _0 (\omega )\frac{{ - i\frac{{\Gamma }}
{2}}} {{\omega  - \Omega   + i\frac{{\Gamma }} {2}}}.
\end{equation}

We note that for Gaussian pulse (\ref{Gauss}) the approximate
probabilities $|\gamma(\omega|^2$ and $|\delta(\omega|^2$ do not
depend on the initial distance $x_0$ between Gaussian peak and the
qubit.

Below in Fig.\ref{Fig2d} and Fig.\ref{Fig2a} we plot the forward,
$S_1(\omega)=|\gamma_{WW}(\omega)|^2\Omega$ and backward,
$S_2(\omega)=|\delta_{WW}(\omega)|^2\Omega$ radiation spectra
calculated from exact expressions (\ref{27a}) and (\ref{28a}), and
systematically compare them with those calculated from
(\ref{31a}), $S_3(\omega)=|\gamma(\omega)|^2\Omega$, and
(\ref{32a}), $S_4(\omega)=|\delta(\omega)|^2\Omega$.

We study how the photon spectra depend on the distance $x_0$ of a
Gaussian peak from qubit. This behavior is shown in
Fig.\ref{Fig2d} for resonance case, $\omega_S=\Omega$ and in
Fig.\ref{Fig2a} and Fig.\ref{Fig3a}  for small detunings,
$\omega_s/\Omega=1.05$ and $\omega_s/\Omega=0.95$, respectively.
For relative large distance, $x_0=0.4$ m $\gg \Delta x$, the exact
equations (\ref{27a}) and (\ref{28a}) provide practically the same
result as the approximate equations (\ref{31a}) and (\ref{32a}).
However, if at the initial instant the Gaussian peak is born
closer to the qubit its front wing begins at the same moment to
interact with a qubit. Therefore, the spectral lines more and more
deviate from large distance results. Finally, we obtain the photon
spectra for $x_0=0$ as shown in the bottom panel of
Fig.\ref{Fig2d}, Fig.\ref{Fig2a}, and Fig.\ref{Fig3a}. We note
that in these figures the quantities
$S_3(\omega)=|\gamma(\omega)|^2\Omega$ and
$S_4(\omega)=|\delta(\omega)|^2\Omega$ calculated from (\ref{31a})
and (\ref{32a}), respectively, do not depend on the distance
between the initial position of Gaussian peak and the qubit.

\begin{figure}
  \includegraphics[width=8 cm]{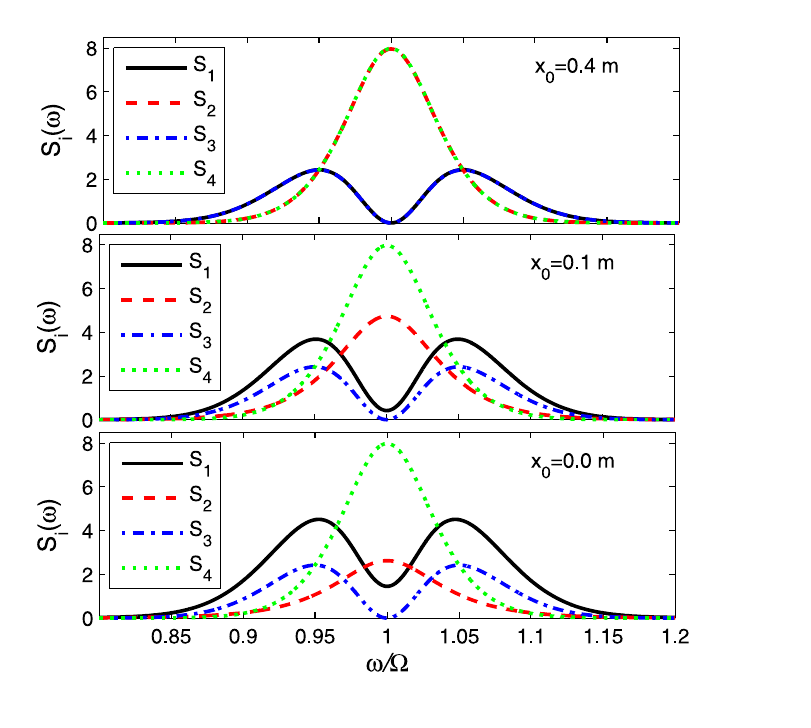}\\
  \caption{Photon radiation spectra for scattering  of Gaussian pulse (\ref{Gauss})
   by a single qubit for different initial distances of the Gaussian peak from the qubit.
$S_1(\omega)=|\gamma_{WW}(\omega)|^2\Omega$ (\ref{27a}), solid
(black) line;
  $S_2(\omega)=|\delta_{WW}(\omega)|^2\Omega$ (\ref{28a}), dashed (red) line;
  $S_3(\omega)=|\gamma(\omega)|^2\Omega$ (\ref{31a}), dashed-dotted (blue) line;
  $S_4(\omega)=|\delta(\omega)|^2\Omega$ (\ref{32a}), dotted (green) line. The parameters of
  the qubit system and initial pulse are as follows:
  $\omega_S/\Omega = 1, \Gamma/\Omega = 0.1,
\Delta/\Omega = 0.1$.}\label{Fig2d}
\end{figure}

\begin{figure}
  \includegraphics[width=8 cm]{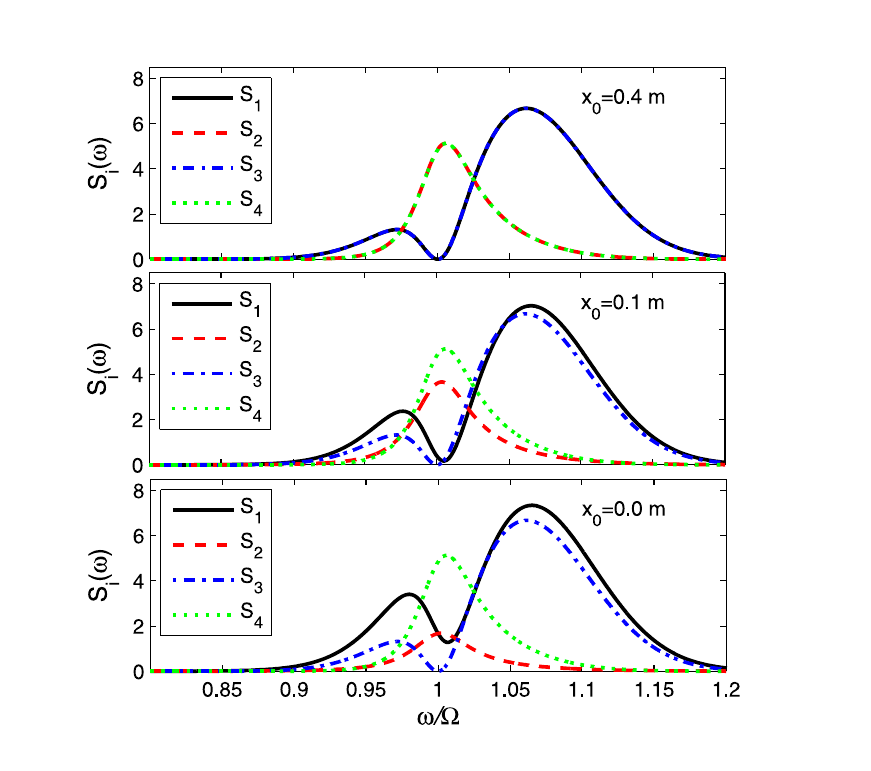}\\
  \caption{Photon radiation spectra for scattering  of Gaussian pulse (\ref{Gauss})
   by a single qubit,
$S_1(\omega)=|\gamma_{WW}(\omega)|^2\Omega$, solid (black) line;
  $S_2(\omega)=|\delta_{WW}(\omega)|^2\Omega$, dashed (red) line;
  $S_3(\omega)=|\gamma(\omega)|^2\Omega$, dashed-dotted (blue) line;
  $S_4(\omega)=|\delta(\omega)|^2\Omega$, dotted (green) line. The parameters of
  the qubit system and initial pulse are as follows:
  $\Delta/\Omega=0.1,\Gamma/\Omega=0.05$, $\omega_s/\Omega=1.05$.}\label{Fig2a}
\end{figure}

\begin{figure}
  \includegraphics[width=8 cm]{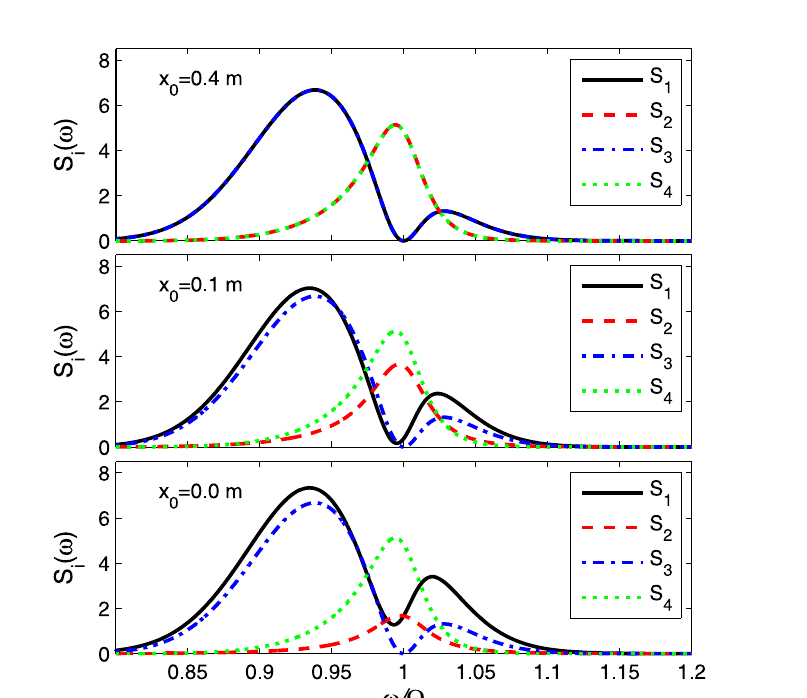}\\
  \caption{Photon radiation spectra for scattering  of Gaussian pulse (\ref{Gauss})
   by a single qubit,
$S_1(\omega)=|\gamma_{WW}(\omega)|^2\Omega$, solid (black) line;
  $S_2(\omega)=|\delta_{WW}(\omega)|^2\Omega$, dashed (red) line;
  $S_3(\omega)=|\gamma(\omega)|^2\Omega$, dashed-dotted (blue) line;
  $S_4(\omega)=|\delta(\omega)|^2\Omega$, dotted (green) line. The parameters of
  the qubit system and initial pulse are as follows:
  $\omega_S/\Omega=1,\Gamma/\Omega=0.05$, $\omega_s/\Omega=0.95$.}\label{Fig3a}
\end{figure}

The equations (\ref{31a}) and (\ref{32a}) provide the flux
conservation at every frequency:
$|\gamma(\omega)|^2+|\delta(\omega)|^2=|\gamma_0(\omega)|^2$.
However, this simple condition is not valid for exact equations
(\ref{27a}) and (\ref{28a}) where the normalization condition has
the form of integral quantity (\ref{4a}). For every plot in
Fig.\ref{Fig2d} and Fig.\ref{Fig2a} we calculated the normalizing
quantity $I = \int {d\omega } \left| {\gamma _{WW} (\omega )}
\right|^2  + \int {d\omega } \left| {\delta _{WW} (\omega )}
\right|^2$ where  the integration was performed within the
frequency span of the plots. In every case, $I$ differs from unity
less than a percent.

The major difference between the plots of equations (\ref{27a}),
(\ref{28a}) and those of equations (\ref{31a}), (\ref{32a}) is
that the transmittance $|\gamma_{WW}(\omega)|^2$ never equals zero
at the resonance frequency, $\omega=\Omega$ and the reflectance
$|\delta_{WW}(\omega)|^2$ never reaches its maximum value
$|\gamma_0(\Omega)|^2$ if the initial distance between Gaussian
peak and the qubit is comparable or less than the pulse width.
These are the principal part integrals in (\ref{27a}), (\ref{28a})
which are responsible for these properties. From the other hand,
the shape of the lines are similar. The transmittance
$|\gamma_{WW}(\omega)|^2$ is up shifted relative to
$|\gamma(\omega)|^2$, while the reflectance
$|\delta_{WW}(\omega)|^2$ is down shifted relative to
$|\gamma(\omega)|^2$.

\section{Two-qubit system}

In this section, we study how a single-photon pulse is scattered
by a two-atom system coupled to a 1D waveguide. We consider two
atoms located at the coordinates $x_1=0$ and $x_2=d$. From
equations (\ref{23}) we obtain two coupled equations for Fourier
components of qubits' amplitudes $\beta_1(\nu)$ and
$\beta_2(\nu)$:
  \begin{subequations}
\begin{align}\label{33}
\begin{gathered}
  \left( {\Delta (\nu ) + i\frac{{\Gamma (\nu )}}
{2}} \right)\beta _1 (\nu ) + i\frac{{\Gamma (\nu )}}
{2}\beta _2 (\nu )\left( {e^{ik_\nu  d}  + iG(k_\nu  d)} \right) \\
   = C_1 (\nu ), \hfill
\end{gathered}
\end{align}
\\
\begin{align}\label{34}
\begin{gathered}
  \left( {\Delta (\nu ) + i\frac{{\Gamma (\nu )}}
{2}} \right)\beta _2 (\nu ) + i\frac{{\Gamma (\nu )}}
{2}\beta _1 (\nu )\left( {e^{ik_\nu  d}  + iG(k_\nu  d)} \right) \hfill \\
   = C_2 (\nu ),  \hfill
\end{gathered}
\end{align}
  \end{subequations}
where $\Delta(\nu)=\nu-\Omega-F(\nu)$, and:
  \begin{subequations}
\begin{align}\label{�1}
C_1 (\nu ) = \pi g(\nu )\gamma _0 (\nu ) + iP\int\limits_0^\infty  {d\omega } \frac{{g(\omega )\gamma _0 (\omega )}}
{{\nu  - \omega }},
\end{align}
\\
\begin{align}\label{�2}
  C_2 (\nu ) = \pi g(\nu )\gamma _0 (\nu )e^{ik_\nu  d}  + iP\int\limits_0^\infty  {d\omega } \frac{{g(\omega )\gamma _0 (\omega )e^{ik_\omega  d} }}
{{\nu  - \omega }}.
\end{align}
  \end{subequations}

From (\ref{33}) we find the explicit expressions for qubits'
amplitudes:
  \begin{subequations}
\begin{align}\label{36}
\begin{gathered}
  \beta _1 (\nu ) = \frac{{\left( {\Delta (\nu ) + i\frac{{\Gamma (\nu )}}
{2}} \right) {C_1 (\nu ) } }} {{\left( {\Delta (\nu ) +
i\frac{{\Gamma (\nu )}} {2}} \right)^2  + \frac{{\Gamma ^2 (\nu
)}}
{4}\left( {e^{ik_\nu  d}  + iG(k_\nu  d)} \right)^2 }} \hfill \\\\
   - \frac{{i\frac{{\Gamma (\nu )}}
{2}\left( {e^{ik_\nu  d}  + iG(k_\nu  d)} \right) {C_2 (\nu ) } }}
{{\left( {\Delta (\nu ) + i\frac{{\Gamma (\nu )}} {2}} \right)^2
+ \frac{{\Gamma ^2 (\nu )}}
{4}\left( {e^{ik_\nu  d}  + iG(k_\nu  d)} \right)^2 }}, \\
\end{gathered}
\end{align}
\\
\begin{align}\label{37}
\begin{gathered}
  \beta _2 (\nu ) = \frac{{\left( {\Delta (\nu ) + i\frac{{\Gamma (\nu )}}
{2}} \right) {C_2 (\nu ) }}} {{\left( {\Delta (\nu ) +
i\frac{{\Gamma (\nu )}} {2}} \right)^2  + \frac{{\Gamma ^2 (\nu
)}}
{4}\left( {e^{ik_\nu  d}  + iG(k_\nu  d)} \right)^2 }} \hfill \\\\
   - \frac{{i\frac{{\Gamma (\nu )}}
{2}\left( {e^{ik_\nu  d}  + iG(k_\nu  d)} \right) {C_1 (\nu ) } }}
{{\left( {\Delta (\nu ) + i\frac{{\Gamma (\nu )}} {2}} \right)^2
+ \frac{{\Gamma ^2 (\nu )}}
{4}\left( {e^{ik_\nu  d}  + iG(k_\nu  d)} \right)^2 }}, \\
\end{gathered}
\end{align}
  \end{subequations}
where:
  \begin{equation}\label{40}
G(k_\nu  d) = \frac{1} {\pi }\cos k_\nu  d\;Ci(k_\nu  d) +
\frac{1} {\pi }\sin k_\nu  d\left( { - \frac{\pi } {2} + Si(k_\nu
d)} \right).
  \end{equation}

From (\ref{24}), (\ref{25}) we obtain the scattering amplitudes
for $t\rightarrow\infty$:

\begin{equation}\label{Tr}
\begin{gathered}
  \gamma (\omega ,t \to \infty )=\gamma _0 (\omega )\\  -i g(\omega )\frac{{\left( {\Delta(\omega)  + \frac{\Gamma }
{2}e^{ - ikd} G(kd)} \right) {C_1 (\omega ) } }} {{\left(
{\Delta(\omega)  + i\frac{\Gamma } {2}} \right)^2 + \frac{{\Gamma
^2 }}
{4}\left( {e^{ikd}  + iG(kd)} \right)^2 }} \hfill \\\\
   -i g(\omega )\frac{{\left( {\Delta(\omega) e^{ - ikd}  + \Gamma \sin kd + \frac{\Gamma }
{2}G(kd)} \right) {C_2 (\omega ) } }} {{\left( {\Delta(\omega)  +
i\frac{\Gamma } {2}} \right)^2  + \frac{{\Gamma ^2 }}
{4}\left( {e^{ikd}  + iG(kd)} \right)^2 }},
\end{gathered}
\end{equation}

\begin{equation}\label{R}
\begin{gathered}
  i\delta (\omega ,t \to \infty )e^{-ikd}/g(\omega )\\ =  \frac{{\left( {\Delta(\omega)  + \frac{\Gamma }
{2}e^{ - ikd} G(kd)} \right) {C_2 (\omega ) )} }} {{\left(
{\Delta(\omega)  + i\frac{\Gamma } {2}} \right)^2 + \frac{{\Gamma
^2 }}
{4}\left( {e^{ikd}  + iG(kd)} \right)^2 }} \hfill \\\\
   + \frac{{\left( {\Delta(\omega) e^{ - ikd}  + \Gamma \sin kd + \frac{\Gamma }
{2}G(kd)} \right) {C_1 (\omega ) }}} {{\left( {\Delta(\omega)  +
i\frac{\Gamma } {2}} \right)^2  + \frac{{\Gamma ^2 }}
{4}\left( {e^{ikd}  + iG(kd)} \right)^2 }},
\end{gathered}
\end{equation}
where  $k=\omega/v_g$, $\Gamma\equiv\Gamma(\omega)$, and
$C_1(\omega)$, $C_2(\omega)$ are given in (\ref{�1}), (\ref{�2}).

We also note here that the derivation of the transmitted and
reflected amplitudes (\ref{Tr}) and (\ref{R}) is not restricted to
Wigner-Weisskopf approximation $\Gamma(\Omega)\ll\Omega$. It is
also valid  for strong coupling where $\Gamma(\Omega)\leq\Omega$.

\subsection{The properties of $G(kd)$}

\begin{figure}
  \includegraphics[width=8 cm]{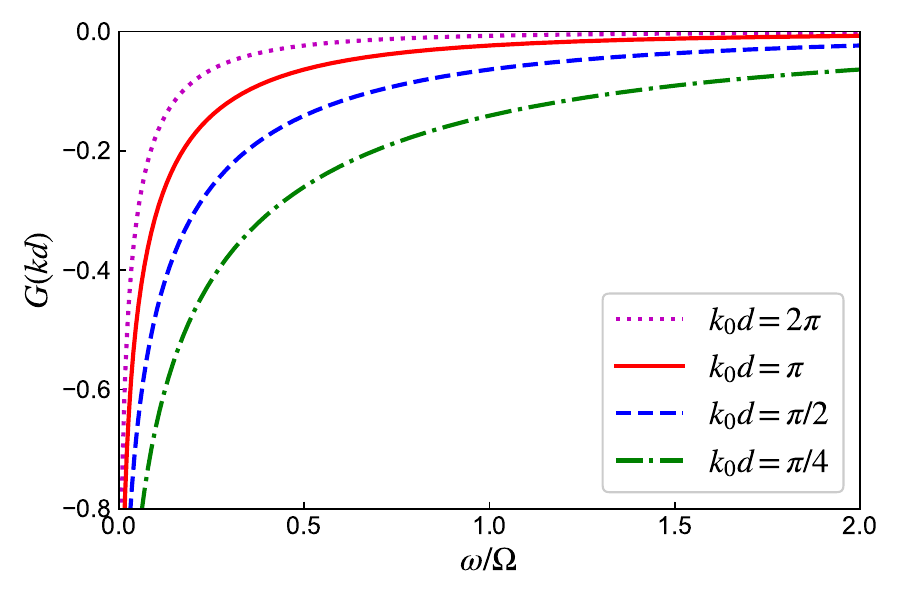}\\
  \caption{The frequency dependence of $G(kd)$ for several interqubit
  distances $d=\lambda,\lambda/2,\lambda/4,\lambda/8$.}\label{Fig1}
\end{figure}

\begin{figure}
  \includegraphics[width=8 cm]{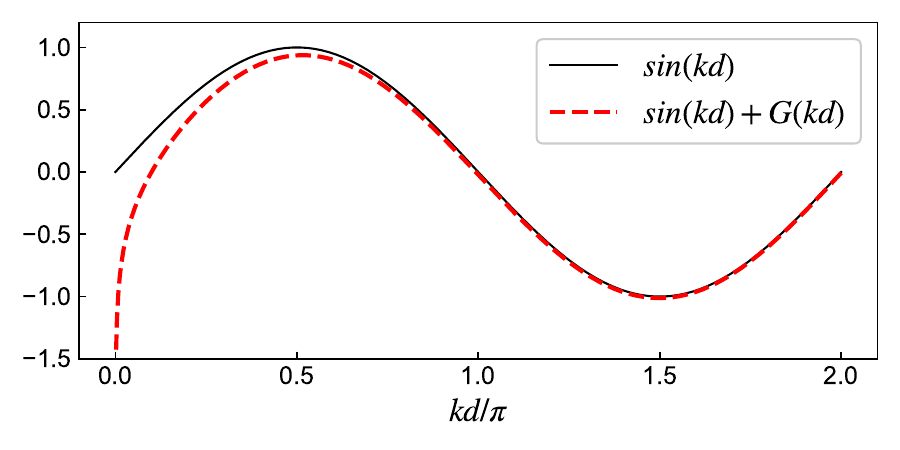}\\
  \caption{The difference between $\sin(kd)$ and $\sin(kd)+G(kd)$ vs. $kd$
  for frequency $\omega=\Omega$.}\label{Fig3}
\end{figure}

\begin{figure}
  \includegraphics[width=8 cm]{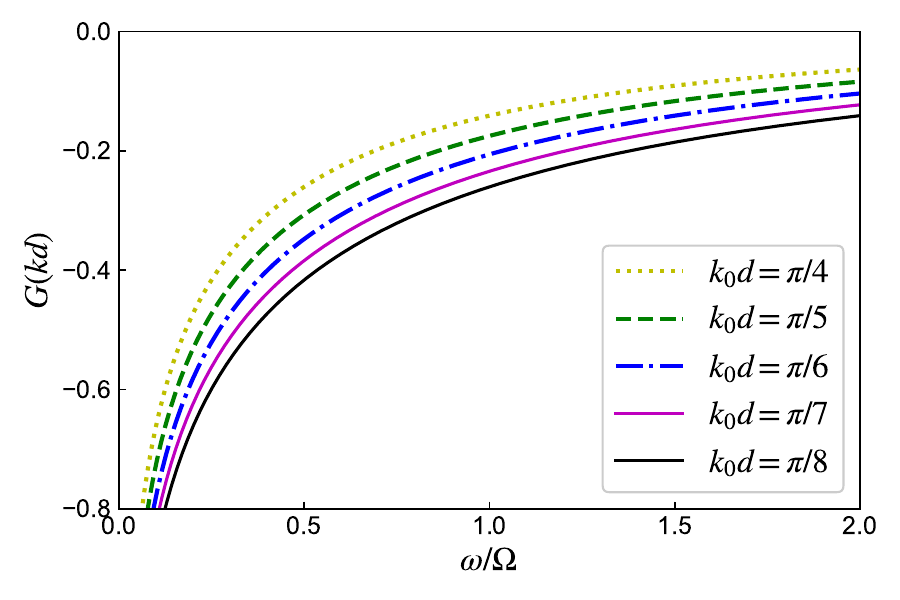}\\
  \caption{The frequency dependence of $G(kd)$ in the range $-0.8<G(kd)<0$
 for several interqubit distances $d=\lambda/8,\lambda/10,\lambda/12,\lambda/14, \lambda/16$.}\label{Fig2}
\end{figure}

From the denominator in (\ref{36}),(\ref{37}) we find the
equations for the resonances (poles) which lie in the lower part
of the complex $\omega$ plane:
\begin{equation}\label{41.3}
\begin{gathered}
  \Delta _ -  (\omega ) =  - \frac{{\Gamma (\omega )}}
{2}\left( {\sin kd + G(kd)} \right) - i\frac{{\Gamma (\omega )}}
{2}\left( {1 - \cos kd} \right), \\
  \Delta _ +  (\omega ) = \frac{{\Gamma (\omega )}}
{2}\left( {\sin kd + G(kd)} \right) - i\frac{{\Gamma (\omega )}}
{2}\left( {1 + \cos kd} \right),
\end{gathered}
\end{equation}
where $k=\omega /v_g$. The numerical calculations show that in a
vast range of relevant parameters the value $G(kd)$ is rather
small (see Fig.\ref{Fig1}). A noticeable difference between
$\sin(kd)$ and $\sin(kd)+G(kd)$ is only observed for $kd<\pi/2$.
(see Fig.\ref{Fig3}). Therefore, for small $k_0d=\Omega d/v_g$ the
main contribution to the frequency shift originates from $G(kd)$
(see Fig.\ref{Fig2}).


The number of poles depends on the interqubit distance $d$.
Assuming Wigner-Weisskopf approximation, $\Gamma(\Omega)/\Omega\ll
1$ we may safely replace in the right hand side of equations
(\ref{41.3}) the running frequency $\omega$ with the qubit
frequency $\Omega$. Thus, in the plane of complex $\omega$ we
obtain two poles:

\begin{equation}\label{42.1}
    \omega_{\pm}=\Omega+\Delta\Omega_{\pm}-i\Gamma_{\pm},
\end{equation}
where $\Delta\Omega_{\pm}$ is the frequency shift:

\begin{equation}\label{42.2}
\Delta \Omega _ \pm   = F(\Omega ) \pm \frac{{\Gamma (\Omega )}}
{2}\left( {\sin k_0 d + G(k_0 d)} \right),
\end{equation}
and $\Gamma_{\pm}$ is the rate of spontaneous emission:

\begin{equation}\label{42.3}
    \Gamma_{\pm}=\frac{\Gamma(\Omega)}{2}\left(1\pm\cos(k_0d)\right),
\end{equation}
where $k_0=\Omega /v_g$.

This two-pole approximation is also called Markovian
approximation. It means that we may neglect the retardation
effects: two qubits feel the incident photon instantaneously. As
was shown in \cite{Zheng2013}  the Markovian approximation is
still valid for $k_0d<5\pi$.

The non-Markovian regime is described by (\ref{41.3}) where
$\omega$ is the running frequency of incident photon.  These
equations imply that the resonance energies  and their widths
depend on the frequency of incident photon, which comes in
(\ref{41.3}) via the the wave vector $k = \omega/v_g$. This is a
general feature of non-Markovian behavior when the photon-
mediated interaction between qubits is not instantaneous and the
retardation effects have to be included. In our method the
retardation effects are automatically included since the quantity
$kd$ explicitly enters the expressions for the transmission and
reflection amplitudes. It is also common to disregard the
frequency dependence of spontaneous emission rate $\Gamma$ keeping
it constant. In this case, non-Markovian behavior manifests via
the the wave vector $k = \omega/v_g$ \cite{Zheng2013}.

From (\ref{42.2}) we see that $G(k_0d)$ is responsible for the
frequency shift, therefore, it modifies the interqubit
interaction. As is seen from (\ref{42.2}) we cannot switch off the
interaction between two qubits taking $k_0d=n\pi$, where $n$ is
integer. However, for the values $k_0d$ for which $\sin(k_0d)=0$
this shift is rather small, $\Delta\Omega_{\pm}/\Gamma(\Omega)\ll
1$. On the other hand, for $k_0d\ll 1$, $G(k_0d)$ scales as
$1/k_0d$. Therefore, in this case, the contribution from $G(k_0d)$
becomes essential. For example, for $k_0d=0.01$, $G(0.01)=-1.28$.
In this case, $\Delta\Omega_{\pm}/\Gamma(\Omega)=\pm 0.64$. This
effect is the evidence of direct dipole-dipole interaction for
which low frequency photons are responsible \cite{Der2014}.

\subsection{Photon amplitudes}

As in the case of a single atom, we can show here that in the
frame of Wigner-Weisskopf approximation and for narrow incident
pulse the equations (\ref{Tr}) and (\ref{R}) provide the
transmitted and reflected amplitudes which are known from the
stationary scattering theories.

The calculation of $C_1(\omega)$ (\ref{�1}) and $C_2(\omega)$
(\ref{�2}) for a delta pulse (\ref{29}) is similar to that of a
single atom (\ref{30}):

\begin{equation}\label{41}
\begin{gathered}
  C_1 (\omega ) = 2\pi g(\omega )\gamma _0 (\omega ), \\
  C_2 (\omega ) = 2\pi g(\omega )\gamma _0 (\omega )e^{ik_\omega  d}. \\
\end{gathered}
\end{equation}

Then, assuming the qubits are initially in the ground state and
using the equations (\ref{24}) and (\ref{25}) we obtain the
transmission and reflection spectra for two-qubit system in the
Wigner-Weisskopf approximation:

  \begin{equation}\label{42}
\begin{gathered}
  \gamma (\omega ,t \to \infty ) \hfill \\
   = \gamma _0 (\omega )\frac{{\Delta _0^2  - \frac{{\Gamma ^2 }}
{4}G^2 (kd) + i\frac{{\Gamma ^2 }} {2}\left( {1 - \cos kd}
\right)G(kd)}} {{\left( {\Delta _0  + i\frac{\Gamma } {2}}
\right)^2  + \frac{{\Gamma ^2 }}
{4}\left( {e^{ikd}  + iG(kd)} \right)^2 }}, \hfill \\
\end{gathered}
  \end{equation}

  \begin{equation}\label{43}
\begin{gathered}
  \delta (\omega ,t \to \infty ) =  - i\frac{\Gamma }
{2}\gamma _0 (\omega )e^{ikd}  \hfill \\
   \times \frac{{2\Delta _0 \cos kd + \Gamma \sin kd + \Gamma G(kd)}}
{{\left( {\Delta _0  + i\frac{\Gamma } {2}} \right)^2  +
\frac{{\Gamma ^2 }}
{4}\left( {e^{ikd}  + iG(kd)} \right)^2 }}, \hfill \\
\end{gathered}
  \end{equation}
where $\Delta_0=\omega-\Omega$, $\gamma_0(\omega)$ is the incident
delta pulse (\ref{29}) and $\Gamma\equiv\Gamma(\Omega)$.

Further simplification can be obtained if the distance between
qubits is sufficiently large, $kd\gg\pi$. For this case, within a
width of the resonance line the quantity $|G(kd)|\ll 1$   (see
Fig.\ref{Fig1}). Hence, we may disregard $G(kd)$ in the
expressions (\ref{42}) and (\ref{43}):

\begin{equation}\label{44}
\gamma (\omega ,t \to \infty ) = \gamma_0(\omega)\frac{{(\omega  -
\Omega )^2 }} {{\left( {\omega  - \Omega  + i\frac{\Gamma } {2}}
\right)^2  + \frac{{\Gamma ^2 }} {4}e^{2ik_\omega  d} }},
\end{equation}

\begin{equation}\label{45}
\begin{gathered}
  \delta (\omega ,t \to \infty ) =  - i\frac{\Gamma }
{2}\gamma _0 (\omega )e^{ik_{\omega}d} \hfill \\
   \times \frac{{2(\omega  - \Omega ) \cos(k_{\omega}d) + \Gamma
\sin(k_{\omega}d)}} {{\left( {\omega  - \Omega  + i\frac{\Gamma }
{2}} \right)^2 + \frac{{\Gamma ^2 }}
{4}e^{2ik_\omega  d} }}. \hfill \\
\end{gathered}
\end{equation}
 The Markovian approximation  is obtained by replacement $k_{\omega}$
with  $k_{\Omega}$  in  (\ref{42}), (\ref{43}), and (\ref{44}),
(\ref{45}).

The expressions (\ref{44}), (\ref{45}) coincide with those
obtained in \cite{Liao2015} for arbitrary shape $\gamma_0(\omega)$
of the incident pulse and with the extension of the coupling to
negative frequencies.

Here we calculate from equations (\ref{Tr}) and (\ref{R}) photon
forward, $\gamma(\omega,t\rightarrow\infty)$ and backward,
$\delta(\omega,t\rightarrow\infty)$ scattering amplitudes. For
simplicity in these equations we assume the rate of spontaneous
emission is constant, $\Gamma(\omega)=\Gamma(\Omega)\equiv\Gamma$.
However, the non-Markovian non-linear dynamics still exists, since
the expressions (\ref{Tr}) and (\ref{R}) depends on photon
frequency via $k=\omega/v_g$. Therefore, the expressions
(\ref{Tr}) and (\ref{R}) transform as follows:

\begin{widetext}
\begin{equation}\label{Tr1}
\begin{gathered}
  \gamma_{WW} (\omega ,t \to \infty ) = \gamma _0 (\omega ) \hfill \\
   -\frac{\Gamma }
{4}\frac{{\left( {\omega  - \Omega  + \frac{\Gamma } {2}e^{ - ikd}
G(kd)} \right)\left(i {\gamma _0 (\omega ) + \frac{1} {\pi
}P\int\limits_0^\infty  {d\omega '\frac{{\gamma _0 (\omega ')}}
{{\omega'  - \omega }}} } \right)}} {{\left( {\omega  - \Omega  +
i\frac{\Gamma } {2}} \right)^2  + \frac{{\Gamma ^2 }}
{4}\left( {e^{ikd}  + iG(kd)} \right)^2 }} \hfill \\
   - \frac{\Gamma }
{4}\frac{{\left( {\left( {\omega  - \Omega } \right)e^{ - ikd}  +
\Gamma \sin kd + \frac{\Gamma } {2}G(kd)} \right)\left( i{\gamma
_0 (\omega )e^{ikd}  + \frac{1} {\pi }P\int\limits_0^\infty
{d\omega '\frac{{\gamma _0 (\omega ')e^{ik'd} }} {{\omega'  -
\omega }}} } \right)}} {{\left( {\omega  - \Omega  + i\frac{\Gamma
} {2}} \right)^2  + \frac{{\Gamma ^2 }}
{4}\left( {e^{ikd}  + iG(kd)} \right)^2 }}, \hfill \\
\end{gathered}
\end{equation}

\begin{equation}\label{R1}
\begin{gathered}
  \delta_{WW} (\omega ,t \to \infty ) = -e^{ikd} \frac{\Gamma }
{4}\frac{{\left( {\omega  - \Omega  + \frac{\Gamma } {2}e^{ - ikd}
G(kd)} \right)\left( {i\gamma _0 (\omega )e^{ikd}  + \frac{1} {\pi
}P\int\limits_0^\infty  {d\omega '} \frac{{\gamma _0 (\omega
')e^{ik'd} }} {{\omega'  - \omega}}} \right)}} {{\left( {\omega -
\Omega  + i\frac{\Gamma } {2}} \right)^2  + \frac{{\Gamma ^2 }}
{4}\left( {e^{ikd}  + iG(kd)} \right)^2 }} \hfill \\
   -e^{ikd} \frac{\Gamma }
{4}\frac{{\left( {(\omega  - \Omega )e^{ - ikd}  + \Gamma \sin kd
+ \frac{\Gamma } {2}G(kd)} \right)\left( i{\gamma _0 (\omega ) +
\frac{1} {\pi }P\int\limits_0^\infty  {d\omega '} \frac{{\gamma _0
(\omega ')}} {{\omega'  - \omega}}} \right)}} {{\left( {\omega -
\Omega  + i\frac{\Gamma } {2}} \right)^2  + \frac{{\Gamma ^2 }}
{4}\left( {e^{ikd}  + iG(kd)} \right)^2 }}. \hfill \\
\end{gathered}
\end{equation}
\end{widetext}

Below we plot the forward and backward radiation spectra
calculated from expressions (\ref{Tr1}) and (\ref{R1}), and
compare them with those calculated from the  expressions
(\ref{44}), (\ref{45}) obtained in \cite{Liao2015}.

\begin{figure}
  \includegraphics[width=8 cm]{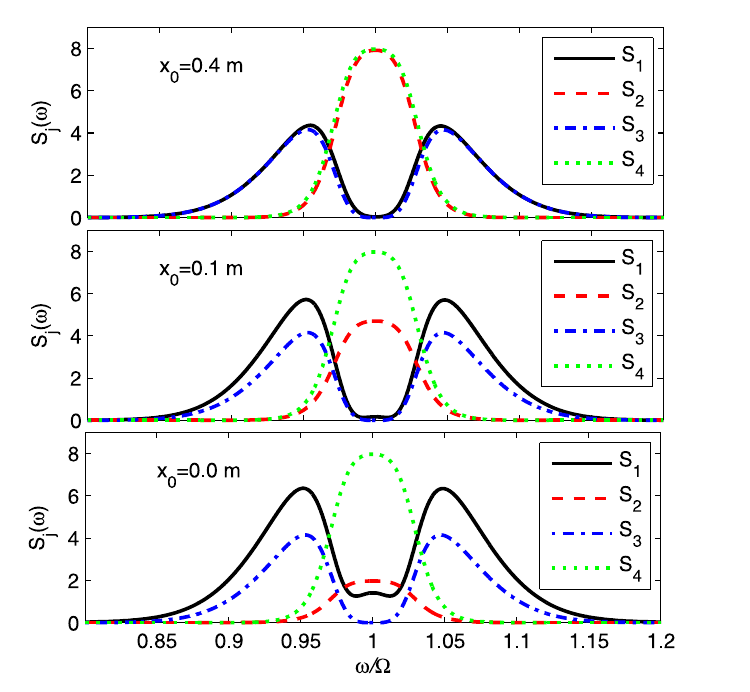}\\
  \caption{Photon radiation spectra for different distances of the Gaussian peak from
  the first qubit for $k_0d=\pi/2$.
$S_1(\omega)=|\gamma_{WW}(\omega)|^2\Omega$, solid (black) line;
  $S_2(\omega)=|\delta_{WW}(\omega)|^2\Omega$, dashed (red) line;
  $S_3(\omega)=|\gamma(\omega)|^2\Omega$, dashed-dotted (blue) line;
  $S_4(\omega)=|\delta(\omega)|^2\Omega$, dotted (green) line. The parameters of
  the qubit system and initial pulse are as follows:
  $\omega_S/\Omega = 1, \Gamma/\Omega = 0.1,
\Delta/\Omega = 0.1$.}\label{Fig14d}
\end{figure}

Similar to the one-qubit case, here the photon spectra also depend
on the distance $x_0$ of a Gaussian peak from qubit. This behavior
is shown in Fig.\ref{Fig14d} for $k_0d=\pi/2$ ($d=\lambda/4$). For
relative large distance, $x_0=0.4$ m $\gg\Delta x$, the exact
equations (\ref{Tr1}) and (\ref{R1}) provide practically the same
results as those of approximate equations (\ref{44}) and
(\ref{45}). As the Gaussian peak becomes closer to the first qubit
the spectral lines more and more deviates from large distance
results. Finally, we obtain the photon spectra for $x_0=0$ as
shown in the bottom panel in Fig.\ref{Fig14d}.

As is known, in N-qubit system the Fano interference gives rise to
sharp asymmetry of spectral lines \cite{Muk2019}. In particular,
the reflected amplitude exhibits $N-1$ zeros which are the
manifestation of the Fano interference. For two-qubit system there
is a single zero of reflected amplitude on the frequency axis, the
position of which depends on qubit parameters, $\Omega$ and
$\Gamma$.

Below we show $x_0$- dependence of transmitted and reflected
amplitudes for the case of Fano interference.

\begin{figure}
   \includegraphics[width=8 cm]{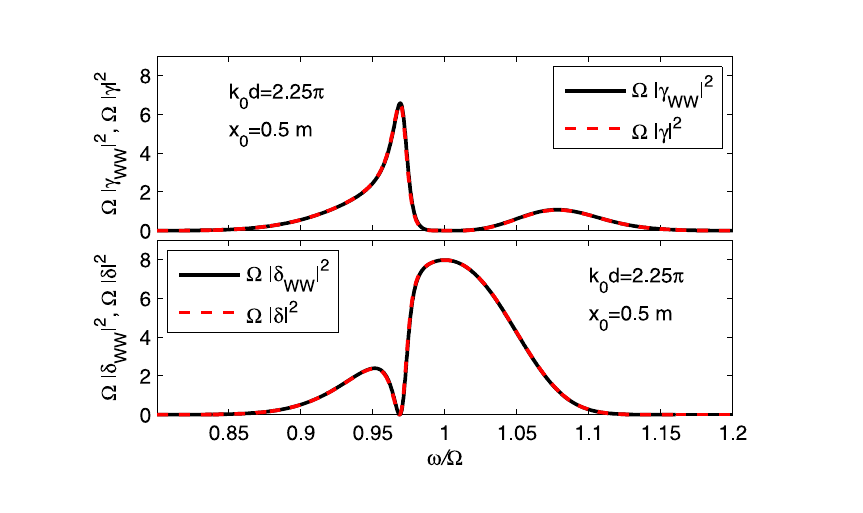}\\
   \caption{Photon radiation spectra of a two-qubit system in
   1D waveguide for incident Gaussian pulse. The transmitted spectra
   $|\gamma_{WW}(\omega,t \rightarrow \infty) |^2 \Omega$ (solid, black line) and
   $|\gamma(\omega,t \rightarrow \infty |^2 \Omega$ (dashed, red
   line) calculated from exact
   (\ref{Tr1}) and approximate (\ref{44}) solutions, respectively, are shown
   in the upper panel. The reflected spectra
    $|\delta_{WW}(\omega,t \rightarrow \infty
|^2 \Omega$ (solid, black line, and $|\delta(\omega,t \rightarrow
\infty) |^2 \Omega$ (dashed, red line) calculated from exact
(\ref{R1}) and approximate (\ref{45})
   solutions, respectively are shown
   in the lower panel.
The parameters of the qubit system and
   initial pulse are:  $k_0d=2.25\pi$ and $x_0=0.5$ m,
   $\omega_S/\Omega = 1,\, \Gamma/\Omega = 0.1,\, \Delta/\Omega = 0.1$.}
   \label{Fig10}
  \end{figure}

\begin{figure}
  \includegraphics[width=8 cm]{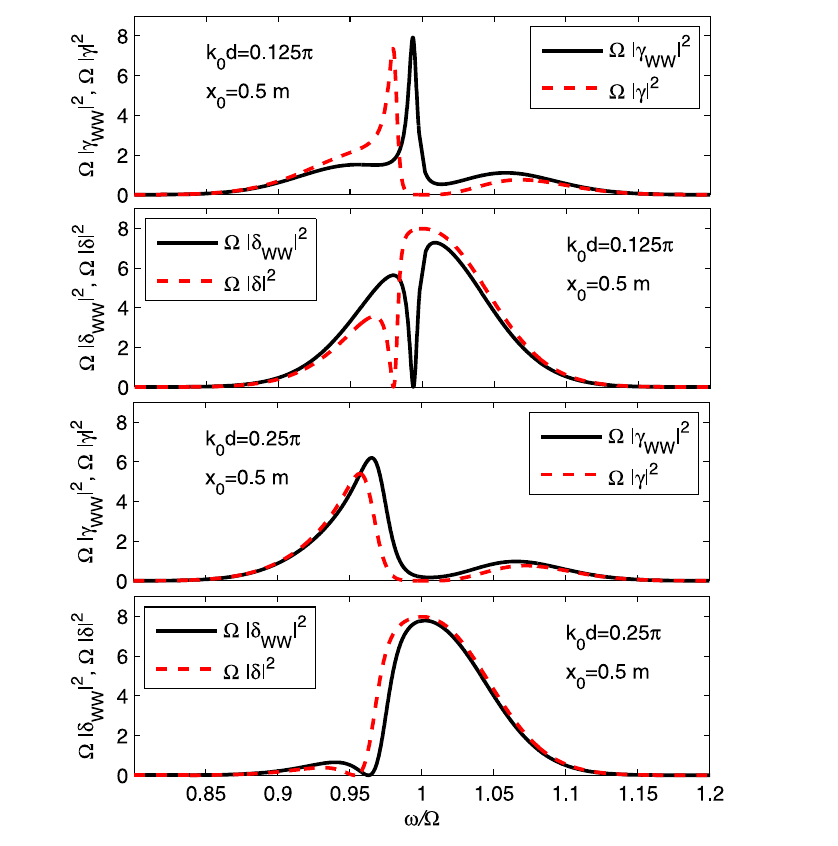}\\
  \caption{Photon radiation spectra of a two-qubit system in
   1D waveguide for incident Gaussian pulse. The transmitted spectra
   $|\gamma_{WW}(\omega,t \rightarrow \infty) |^2 \Omega$ (solid, black line) and
   $|\gamma(\omega,t \rightarrow \infty |^2 \Omega$ (dashed, red
   line) calculated from exact
   (\ref{Tr1}) and approximate (\ref{44}) solutions, respectively, are shown
   in the upper  two panels. The reflected spectra
    $|\delta_{WW}(\omega,t \rightarrow \infty
|^2 \Omega$ (solid, black line, and $|\delta(\omega,t \rightarrow
\infty) |^2 \Omega$ (dashed, red line) calculated from exact
(\ref{R1}) and approximate (\ref{45})
   solutions, respectively are shown
   in the lower two panels.
The parameters of the qubit system and
   initial pulse are:  $x_0=0.5$ m,
   $\omega_S/\Omega = 1,\, \Gamma/\Omega = 0.1,\,
   \Delta/\Omega = 0.1$.}\label{Fig10a}
\end{figure}

The transmitted and reflected spectra for $k_0d = 2.25\pi$, and
$x_0=0.5$ m are shown in Fig. \ref{Fig10}. We see that the exact
spectra (\ref{Tr1}),(\ref{R1})  coincide with those calculated
from approximate expressions (\ref{44}), (\ref{45}). There  are
two reasons for this. First, the distance between the initial
position of Gaussian peak and the first qubit is  much larger than
the pulse width ($50$ cm and $10$ cm, respectively). Second, the
influence of $G(kd)$ on spectral lines for $k_0d = 2.25\pi$ is
rather small within the  width of line resonances (see
Fig.\ref{Fig1}). A signature of Fano interference is seen not only
for reflected spectrum (zero in the lower plot in Fig.
\ref{Fig10}) but also for transmitted spectrum (sharp asymmetrical
shape of transmitted line in the upper plot in Fig. \ref{Fig10}).

The influence of $G(kd)$ on transmitted and reflected spectra is
shown in Fig. \ref{Fig10a}. Here $x_0$
 is sufficiently large, therefore, the deviation from approximate
 spectral lines can be attributed to $G(kd)$ in exact expressions.
 The contribution of $G(kd)$ can be seen by comparison the upper
 two panels with lower two panels in Fig. \ref{Fig10a}. The
 smaller $k_0d$ the larger $G(kd)$ within the width of the spectral lines.
 Therefore, for $k_0d=0.125\pi$ the deviation of exact resonance lines from
 approximate ones is larger than that for $k_0d=0.25\pi$.

  \begin{figure}
   \includegraphics[width=8 cm]{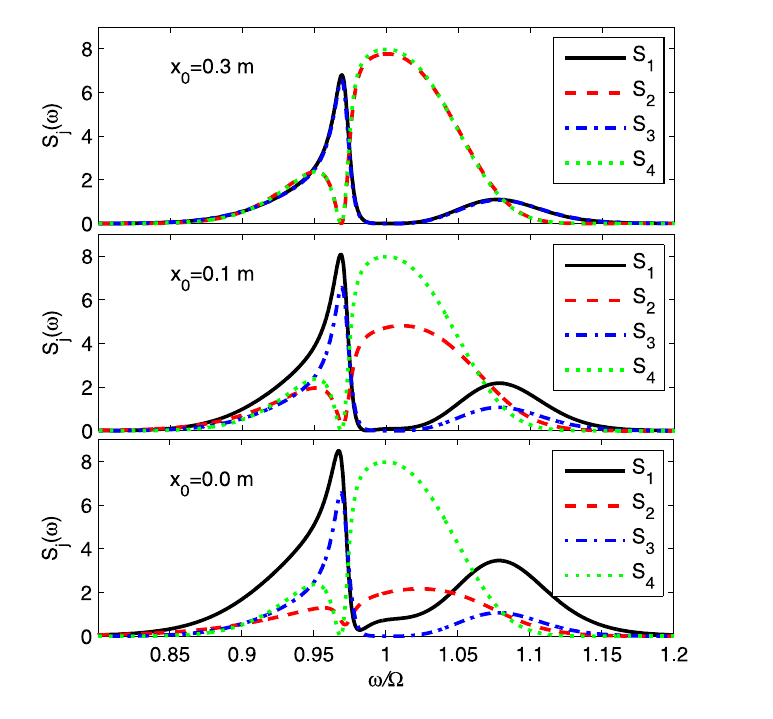}\\
   \caption{Photon radiation spectra of a two-qubits for different
   distance $x_0$ between the initial position of the pulse peak
   and the first qubit.
   All plots show both
   transmitted ($S_1=|\gamma_{WW}(\omega,t \rightarrow \infty) |^2 \Omega$, $S_3=
   |\gamma(\omega,t \rightarrow \infty |^2 \Omega$) and reflected ($S_2=
   |\delta_{WW}(\omega,t \rightarrow \infty) |^2 \Omega$, $S_4=
   |\delta(\omega,t \rightarrow \infty) |^2 \Omega$) radiation spectra.
   Solid (black) and dashed (red) lines represent the exact solutions
   (\ref{Tr1}) and (\ref{R1}),
  dotted-dashed (blue) and dotted (green) lines represent the approximate solutions (\ref{44}), (\ref{45}), The parameters of the qubit system and
   initial pulse are:
    $\omega_S/\Omega = 1,\, \Delta/\Omega = 0.1,\, \Gamma/\Omega = 0.1,\, k_0d = 2.25\pi$.} \label{Fig11}
  \end{figure}
The dependence of transmitted and reflected spectra on the initial
distance, $x_0$ between a Gaussian peak and the first qubit is
shown in Fig.\ref{Fig11}. These plots are calculated for
$k_0d=2.25\pi$ which makes it possible to neglect the contribution
from $G(kd)$ in exact equations (\ref{Tr1}) and (\ref{R1}) leaving
the bare $x_0$-dependence. The spectral amplitudes calculated from
approximate expressions (\ref{44}) and (\ref{45}) do not depend on
$x_0$. If $x_0$ is large compared to the pulse width the exact
spectral amplitudes are close to the approximate ones (the upper
panel in Fig.\ref{Fig11}). The maximum deviation we see for
$x_0=0$ (the lower panel in Fig.\ref{Fig11}).

\begin{figure}
   \includegraphics[width=8 cm]{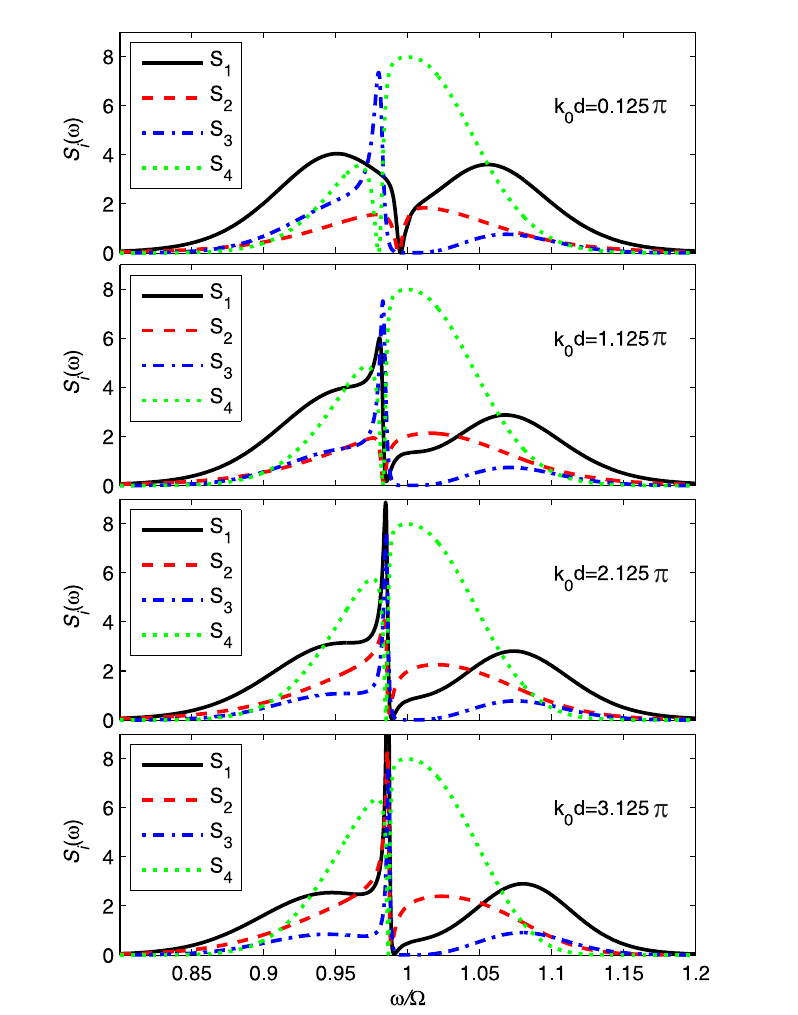}\\
   \caption{Photon radiation spectra of a two-qubit system
 for different values of $k_0d$. All plots show both
   transmitted ($S_1=|\gamma_{WW}(\omega,t \rightarrow \infty) |^2 \Omega$, $S_3=
   |\gamma(\omega,t \rightarrow \infty |^2 \Omega$) and reflected ($S_2=
   |\delta_{WW}(\omega,t \rightarrow \infty) |^2 \Omega$, $S_4=
   |\delta(\omega,t \rightarrow \infty) |^2 \Omega$) radiation spectra.
   Solid (black) and dashed (red) lines represent the exact solutions
   (\ref{Tr1}) and (\ref{R1}),
  dotted-dashed (blue) and dotted (green) lines represent the approximate
  solutions (\ref{44}), (\ref{45}).
 The parameters of the qubit
 system and initial pulse are:
  $\Gamma/\Omega = 0.1,\Delta/\Omega = 0.1, \omega_S=\Omega,\,x_0=0$.} \label{Fig14}
  \end{figure}

Finally, we show in Fig.\ref{Fig14} the spectral amplitudes for
different values, $k_0d$ calculated for $x_0=0$.  The radiation
spectra are shown in Fig.\ref{Fig14} for $k_0d=0.125\pi
\; (d=\lambda/16)$, $k_0d=1.125\pi \; (d=9\lambda/16)$, $k_0d=2.125\pi
\; (d=17\lambda/16)$, $k_0d=3.125\pi \; (d=25\lambda/16)$. The different
forms of spectral lines, calculated from exact expressions are
only due to the contribution from $G(kd)$. Compare upper panel in
Fig.\ref{Fig14} where $k_0d=0.125\pi$ with the lower panel where
$k_0d=3.125\pi$.

It also worth noting very sharp and narrow peaks for transmission
radiation $S_1$. The peak at the bottom panel in Fig.\ref{Fig14}
($k_0d=3.125\pi$) has its maximum at 14.7 which is not shown in
the panel. These narrow peaks are clear signature of the effects
of non-Markovianity and subradiant transitions, which yields  the
ultra narrow emission lines.

For every plot in Fig.\ref{Fig14d}-Fig.\ref{Fig14} we controlled
the normalizing quantity $I = \int {d\omega } \left| {\gamma _{WW}
(\omega )} \right|^2  + \int {d\omega } \left| {\delta _{WW}
(\omega )} \right|^2$ where the integration was performed within
the frequency span of the plots. In every case, $I$ differed from
unity less than several thousandths.

\section{Conclusion}
In this paper we analyze the scattering of a single-photon pulse
by a one-dimensional chain of two level artificial atoms (qubits)
embedded in an open waveguide. The system is described by the
continuum mode Jaynes-Cummings Hamiltonian with Hilbert space
being truncated to a single excitation subspace. The
time-dependent dynamical equations for qubits' amplitudes and for
transmitted and reflected photon spectra are solved by application
of Heitler's method. Unlike previous approaches our calculations
are performed only for physical, positive, frequency axis. This
significantly changes the dynamics of the system. First, it leads
to the additional photon-mediated dipole-dipole interaction
between qubits which results in the violation of the phase
coherence between them. Second, the spectral lines of transmitted
and reflected spectra crucially depend on the shape of incident
pulse. We apply our theory to one-qubit and two-qubit systems. For
these two cases we obtain the explicit expressions for the qubits'
amplitudes and for the photon spectra. For the travelling incident
Gaussian wave packet we calculate the line shapes of transmitted
and reflected photons. We show that the transmitted and reflected
photon spectra crucially depend on the initial distance of the
Gaussian peak from the first qubit. If a distance between the
initial position of a pulse peak is comparable to or less than the
pulse width in a configuration space the spectral lines are
significantly modified. As this distance becomes closer to the
qubit, the spectral lines more and more deviate from known results
where the continuation of the qubit-photon interaction to the
negative frequency axis is used. For two-qubit radiation spectra
we thoroughly investigate the influence of $G(kd)$ and position of
the pulse peak on Fano interference which gives rise to a sharp
asymmetrical form of the the spectral lines.

We believe that the results obtained in the paper may be important
for the practical implementation in waveguide QED problems with
superconducting qubits where the control and readout pulse
generators should be placed as close as possible to the qubit chip
at millikelvin temperatures \cite{Lec2021}.

Our approach can also be further extended and applied to the
single-photon scattering by multi-level qubits in strong
light-matter coupling regimes.

\begin{acknowledgments}

The authors thank O. V. Kibis  and A. N. Sultanov for fruitful
discussions. The work is supported by the Ministry of Science and
Higher Education of Russian Federation under the project
FSUN-2023-0006. O. Chuikin acknowledges the financial support from
the Foundation for the Advancement of Theoretical Physics and
Mathematics “BASIS”.
\end{acknowledgments}

\appendix
\section{Derivation of equation (\ref{G})}

We begin with the integral in equation (\ref{22}):
\begin{equation}\label{A1}
I(\nu)  = \int\limits_0^\infty  {} d\omega g^2 (\omega )\cos
\left( {k(x_n  - x_{n'} )} \right)\zeta (\nu  - \omega ).
\end{equation}
Using (\ref{21}) we obtain:
\begin{equation}\label{A2}
\begin{gathered}
  I(\nu ) =  - i\pi g^2 (\nu )\cos k_\nu  d_{nn'}  + I_ -  (\nu ) \hfill \\
   =  - i\pi g^2 (\nu )e^{ik_\nu  d_{nn'} }  + \left( {I_ -  (\nu ) - \pi g^2 (\nu )\sin k_\nu  d_{nn'} } \right), \hfill \\
\end{gathered}
\end{equation}
where:
\begin{equation}\label{A3}
    I_-(\nu)= -P\int\limits_0^\infty  {} d\omega \frac{{g^2 (\omega )\cos
\left( {kd_{nn'} } \right)}} {{\omega-\nu  }}.
\end{equation}

Usually, this integral is calculated by taking $g^2(\omega)$ out
of the integral at the frequency $\nu$ and moving the lower bound
to $-\infty$:
\begin{equation}\label{AA4}
I_ -  (\nu ) \approx  - g^2 (\nu )P\int\limits_{ - \infty }^\infty
{d\omega } \frac{{\cos \left( {\omega t_{nn'} } \right)}} {{\omega
- \nu }} = \pi g^2 (\nu )\sin \left( {\nu t_{nn'} } \right),
\end{equation}
where  $t_{nn'}=d_{nn'}/v_g$ and we used Kramers-Kronig relation.
In this case the second term in brackets in (\ref{A2}) is zero.
Therefore, for $I(\nu)$ we obtain: $I(\nu)=-i\pi g^2(\nu)\exp(i\nu
t_{nn'})$.

We would obtain (\ref{AA4}) if we added to $I_-(\nu)$ (\ref{A3})
the non-RWA counter rotating term $I_+(\nu)$:

\begin{equation}\label{AA5}
    I_+(\nu)= -P\int\limits_0^\infty  {} d\omega \frac{{g^2 (\omega )\cos
\left( {kd_{nn'} } \right)}} {{\omega+\nu  }}.
\end{equation}

If we assume $g^2(\omega)=\lambda\omega$, change variables in the
integrand in (\ref{AA5}) ($\omega\rightarrow -\omega$), and add
the result to $I_-(\nu)$ we obtain the result which is given in
(\ref{AA4}). Strictly speaking, this procedure is not well
justified because any frequency dependent coupling ($g^2(\omega)$
in our case) must be exactly zero for negative frequencies.

In order to estimate the contribution from non RWA term more
elaborate calculation of $I_-(\nu)$ (\ref{A3}) is necessary.
First, we add and subtract to $I_-(\nu)$ the counter rotating term
$I_+(\nu)$: $I_-(\nu)=[I_-(\nu)+I_+(\nu)]-I_+(\nu)=\pi
g^2(\nu)\sin(\nu t_{nn'})-I_+(\nu)$. Therefore, for the second
term in right hand side in (\ref{A2}) we obtain:

\begin{equation}\label{AA6}
I_+(\nu)=-\left( {I_ -  (\nu ) - \pi g^2 (\nu )\sin k_\nu  d_{nn'}
} \right).
\end{equation}

Next, we begin calculating the quantity $I_+(\nu)$.
We assume $g^2(\omega)=\lambda\omega$. Then, for $I_+(\nu)$ we
obtain:
\begin{equation}\label{A4}
\begin{gathered}
  I_+ (\nu ) =  - \lambda P\int\limits_0^\infty  {} d\omega \frac{{\omega \cos \left( {\omega t_{nn'} } \right)}}
{{\omega  + \nu }} \hfill \\
   =  - \lambda \int\limits_0^\infty  {} d\omega \cos \left( {\omega t_{nn'} } \right) +\lambda \nu P\int\limits_0^\infty  {} d\omega \frac{{\cos \left( {\omega t_{nn'} } \right)}}
{{\omega  + \nu }}. \hfill \\
\end{gathered}
\end{equation}

To calculate the first term of the right-hand side, we add a small
converging factor. This reflects the fact that the system does not
respond at high frequencies. In this way, we find that this term
vanishes:

\begin{equation}\label{A5}
\begin{gathered}
  \int\limits_0^\infty  {} d\omega \cos \left( {\omega t_{nn'} } \right) = \nu \int\limits_0^\infty  {} dy\cos \left( {y\nu t_{nn'} } \right) \hfill \\
   = \nu \mathop {\lim }\limits_{\eta  \to 0^ +  } \int\limits_0^\infty  {} dye^{ - \eta y} \cos y\nu t_{nn'}  = \nu \mathop {\lim }\limits_{\eta  \to 0^ +  } \frac{\eta }
{{(\nu t_{nn'} )^2  + \eta ^2 }} = 0. \hfill \\
\end{gathered}
\end{equation}

In the integrand of the second term we change variable
$x=(\omega+\nu)/\nu$:
\begin{equation}\label{A6}
\begin{gathered}
  P\int\limits_0^\infty  {} d\omega \frac{{\cos \left( {\omega t_{nn'} } \right)}}
{{\omega  + \nu }} = P\int\limits_{  1}^\infty  {} dx\frac{{\cos
\left( {(x - 1)\nu t_{nn'} } \right)}}
{x} \hfill \\
   = \cos \nu t_{nn'} P\int\limits_{  1}^\infty  {} dx\frac{{\cos \left( {x\nu t_{nn'} } \right)}}
{x}\\ + \sin \nu t_{nn'} P\int\limits_{  1}^\infty  {}
dx\frac{{\sin \left( {x\nu t_{nn'} } \right)}}
{x}. \\
\end{gathered}
\end{equation}
Two integrals in (\ref{A6}) can be expressed in terms of sine and
cosine integral functions which are given in (\ref{CS}).
Therefore, for the quantity $I_+(\nu)$ we obtain:

\begin{equation}\label{A7}
\begin{gathered}
  I_+ (\nu ) = -g^2 (\nu )\cos \nu t_{nn'} Ci(\nu |t_{nn'}| ) \\
   + g^2 (\nu )\sin \nu t_{nn'} \left( {\frac{\pi }
{2}\operatorname{sgn} ( t_{nn'} ) - Si(\nu t_{nn'} )} \right).
\end{gathered}
\end{equation}

As is seen from (\ref{A7}), the quantity $I_+(\nu)$ is the even
function of $t_{nn'}$. Therefore we may rewrite (\ref{A7}) in the
following form:
\begin{equation}\label{A8}
\begin{gathered}
  I_+ (\nu ) = -g^2 (\nu )\cos \nu t_{nn'} Ci(\nu |t_{nn'}| ) \\
   + g^2 (\nu )\sin \nu \left| {t_{nn'} } \right|\left( {\frac{\pi }
{2} - Si(\nu \left| {t_{nn'} } \right|)} \right).
\end{gathered}
\end{equation}

Finally, for $I(\nu)$ we obtain:
\begin{equation}\label{A9}
\begin{gathered}
  I (\nu ) =  - i\frac{{\Gamma (\nu )}}
{4}e^{i\nu \left| {t_{nn'} } \right|}  + \frac{{\Gamma (\nu )}}
{{4\pi }}\cos \nu t_{nn'} Ci(\nu |t_{nn'}| )  \\
   + \frac{{\Gamma (\nu )}}
{{4\pi }}\sin \nu \left| {t_{nn'} } \right|\left( { - \frac{\pi }
{2} + Si(\nu \left| {t_{nn'} } \right|)} \right).
\end{gathered}
\end{equation}
The insertion of $I(\nu)$ in (\ref{22}) provides the equation
(\ref{23}) with the quantity $G(kd_{nn'})$ defined in (\ref{G}).

In fact, according to the derivation the quantity $G(kd_{nn'})$
can be expressed as:
\begin{equation}\label{A10}
G(k_\nu  d_{nn'} ) =  - \frac{4} {{\Gamma (\nu )}}I_ +  (\nu ) = -
\frac{4} {{\Gamma (\nu )}}\int\limits_0^\infty  {d\omega
\frac{{g^2 (\omega )\cos k_\nu  d_{nn'} }} {{\omega  + \nu }}}.
\end{equation}
Simply speaking, $G(kd_{nn'})$ is the difference between $\pi
g^2(\nu)\sin(\nu t_{nn'})$ which accounts for the non RWA
contribution ($\pi g^2(\nu)\sin(\nu t_{nn'})=I_-(\nu)+I_+(\nu)$)
and $I_-(\nu)$:
\begin{equation}\label{A11}
G(k_\nu  d_{nn'} ) =  - \frac{4} {{\Gamma (\nu )}}\left( {\pi g^2
(\nu )\sin (\nu t_{nn'} ) - I_ -  (\nu )} \right).
\end{equation}

The numerical calculations of $G(kd)$ which are presented in Figs.
\ref{Fig1}, \ref{Fig2}, \ref{Fig3} in the main text show that
within resonance width the non-RWA contribution is rather small
for $kd_{nn'}>\pi/4$. However, this is not the case if
$kd_{nn'}<\pi/4$.

\section{Proof of equation (\ref{bt0})}

In the Wigner-Weisskopf approximation $\Gamma(\Omega)\ll\Omega$.
Therefore, in (\ref{bt}) we may ignore $F(\omega)$ and replace
$\Gamma(\omega)$ with $\Gamma(\Omega)\equiv\Gamma$. Further, we
use for $\zeta (\omega-\omega')$ one of its representation:
  \begin{equation}\label{B1}
\zeta (\omega  - \omega' ) = \mathop {\lim }\limits_{\sigma  \to
0} \frac{1} {{\omega  - \omega'  + i\sigma }}.
  \end{equation}
Therefore, for (\ref{bt}) we obtain:
\begin{equation}\label{B2}
\begin{gathered}
  \beta (t) = i \int\limits_0^\infty  {d\omega 'g(\omega ')\gamma _0 (\omega ')}  \hfill \\
   \times \mathop {\lim }\limits_{\sigma  \to 0}\int\limits_{ - \infty }^\infty  {\frac{{d\omega }}
{{2\pi }}} \frac{{e^{ - i(\omega  - \Omega )t} }} {{\left( {\omega
- \Omega  + i\frac{\Gamma }
{2}} \right)\left( {\omega  - \omega ' + i\sigma } \right)}}. \hfill \\
\end{gathered}
\end{equation}
There are only two poles in the lower part of the complex $\omega$
plane. Applying the Cauchy residue theorem to the second integral
in (\ref{B2}) we obtain the following result:
\begin{equation}\label{B3}
\beta (t) = \int\limits_0^\infty  {d\omega 'g(\omega ')\gamma _0
(\omega ')} \frac{{e^{ - \frac{\Gamma } {2}t}  - e^{ - i(\omega '
- \Omega )t} }} {{\left( {\omega ' - \Omega  + i\frac{\Gamma }
{2}} \right)}},
\end{equation}
for $t>0$ and $\beta (t)=0$ for $t<0$.

\end{document}